\documentclass[14pt]{ucthesis}

\makeatletter
\renewcommand{\normalsize}{\@setfontsize\normalsize\@xipt{13.6}%
\abovedisplayskip 11\p@ plus3\p@ minus6\p@
\belowdisplayskip \abovedisplayskip
\abovedisplayshortskip  \z@ plus3\p@  
\belowdisplayshortskip  6.5\p@ plus3.5\p@ minus3\p@
\let\@listi\@listI}   
\renewcommand{\small}{\@setfontsize\small\@xpt{12}%
\abovedisplayskip 10\p@ plus2\p@ minus5\p@
\belowdisplayskip \abovedisplayskip
\abovedisplayshortskip  \z@ plus3\p@  
\belowdisplayshortskip  6\p@ plus3\p@ minus3\p@
\def\@listi{\leftmargin\leftmargini 
\topsep 6\p@ plus2\p@ minus2\p@\parsep 3\p@ plus2\p@ minus\p@
\itemsep \parsep}}
\renewcommand{\footnotesize}{\@setfontsize\footnotesize\@ixpt{11}%
\abovedisplayskip 8\p@ plus2\p@ minus4\p@
\belowdisplayskip \abovedisplayskip
\abovedisplayshortskip \z@ plus\p@
\belowdisplayshortskip 4\p@ plus2\p@ minus2\p@
\def\@listi{\leftmargin\leftmargini 
\topsep 4\p@ plus2\p@ minus2\p@\parsep 2\p@ plus\p@ minus\p@
\itemsep \parsep}}
\renewcommand{\scriptsize}{\@setfontsize\scriptsize\@viiipt{9.5pt}}
\renewcommand{\tiny}{\@setfontsize\tiny\@vipt{7pt}}
\renewcommand{\large}{\@setfontsize\large\@xiipt{16pt}}
\renewcommand{\Large}{\@setfontsize\Large\@xivpt{18pt}}
\renewcommand{\LARGE}{\@setfontsize\LARGE\@xviipt{22pt}}
\renewcommand{\huge}{\@setfontsize\huge\@xxpt{25pt}}
\renewcommand{\Huge}{\@setfontsize\Huge\@xxvpt{30pt}}
\makeatother
\usepackage{epsfig}
\usepackage{times}
\usepackage{amsmath,amsfonts,amssymb,graphicx,graphics}
\usepackage{cite}
\usepackage[varg]{txfonts}
\usepackage{array}
\usepackage{multirow}
\usepackage{rotating}
\usepackage{setspace}
\usepackage{color}
\usepackage{caption}
\usepackage{subfig}
\usepackage{soul,color}
\usepackage{hhline}
\usepackage{footnote}
\usepackage[bottom]{footmisc}
\usepackage{setspace}
\usepackage{morefloats}
\usepackage{bibentry}
\usepackage{multicol}
\usepackage{fancyhdr}
\usepackage{algorithm}
\usepackage{algorithmic}
\usepackage{placeins}
\usepackage{url}
\usepackage{paralist}
\usepackage{comment}


\makeatother

\ssp

\newcommand{\eat}[1]{}

\makeatletter
 {\vskip 5pt\begin{list}{}{%
  \setlength{\topsep}{0pt}\setlength{\partopsep}{0pt}\setlength{\parskip}{0pt}%
  \setlength{\parsep}{0pt}\setlength{\labelwidth}{0pt}%
  \setlength{\rightmargin}{0pt}\setlength{\leftmargin}{0pt}%
  \setlength{\labelsep}{0pt}%
  \obeylines\@vobeyspaces\normalfont\ttfamily%
  \item[]}}
 {\end{list}\vskip5pt\noindent}
\makeatother

\begin{document}


\title{High Throughput Push Based Storage Manager}
\author{Ye Zhu}
\degreeyear{2019}
\degreesemester{Spring}
\degree{Master in Computer Science}

\chair{Professor Florin Rusu}
\othermembers{Professor Mukesh Singhal\\
Professor Dong Li}
\chairname{Florin Rusu}
\othermembera{Mukesh Singhal}
\othermemberb{Dong Li}
\numberofmembers{3}
\prevdegrees{B.E. (Beijing Institute of Technology, Beijing, China) 2010\\
PHD. (University of California, Merced, CA, USA) 2016}
\field{Electronic Engineering and Computer Science}
\campus{Merced}

\maketitle
\copyrightpage
\approvalpage

\begin{frontmatter} 
\begin{acknowledgements}
    
\vspace{28pt}
First of all, I would like to sincerely thank my advisor Professor Florin Rusu. Without his constant guidance over the years, this project would not have been possible. I was inspired and enlightened by his profound knowledge, acute insights, deep understanding of the philosophy of research, sense of humor and most importantly, meticulous attention to details.
    
\vspace{18pt}
I would like to thank my dissertation committee, Professor Mukesh Singhal and Professor Dong Li from Electronic Engineering and Computer Science for their time, insightful comments and constructive guidance that significantly improved this dissertation.
    
\vspace{18pt}
I would like to express my appreciation to my labmates and colleagues at UC Merced. Yu Cheng and Xin Zhang offered me valuable advice when I began my research on this dissertation project. Also the labmates Weijie Zhao, Yujing Ma, Zhiyi Huang, and Jun. 
    
\vspace{18pt}
Thanks to my parents and my family. Without their support and love, I won't be able to accomplish this dissertation.
    
\vspace{18pt}
I  would like to thank the numerous fellowship awards from UC Merced, financial support and training from being research and teaching assistant at UC Merced.

\end{acknowledgements}

\tableofcontents
\listoffigures

\end{frontmatter}

\begin{abstract} 
	\begin{large}
The storage manager, as a key component of the database system, is responsible for organizing, reading, and delivering data to the execution engine for processing. According to the data serving mechanism, existing storage managers are either pull-based, incurring high latency, or push-based, leading to a high number of I/O requests when the CPU is busy. To improve these shortcomings, this thesis proposes a push-based prefetching strategy in a column-wise storage manager. The proposed strategy implements an efficient cache layer to store shared data among queries to reduce the number of I/O requests. The capacity of the cache is maintained by a time access-aware eviction mechanism. Our strategy enables the storage manager to coordinate multiple queries by merging their requests and dynamically generate an optimal read order that maximizes the overall I/O throughput. We evaluated our storage manager both over a disk-based redundant array of independent disks (RAID) and an NVM Express (NVMe) solid-state drive (SSD). With the high read performance of the SSD, we successfully minimized the total read time and number of I/O accesses. 
	\end{large}
\end{abstract}

\chapter{Introduction}

\section{Background and Motivation}
\begin{large}
In the era of data deluge, massive quantities of data are generated at an unprecedented scale by applications ranging from social networks to scientific experiments and personalized medicine. The vast majority of this read-only data is stored as application-specific files containing hundreds of millions of records.

The first step is to understand the characteristics of database workloads running on the system. According to the book~\cite{dbreadings}, the database systems can be roughly divided into two categories: on-line transaction processing (OLTP) and on-line analytical processing (OLAP).

OLTP is a class of application that manages data entry retrieval transactions in many industries, including banking, supermarkets, airlines, rental services, and manufacturing. A typical OLTP workload contains simple transactions from many terminals with strict response time requirements (several seconds). Many of these transactions update the database. In OLTP workloads, traffic to disks is characterized by small random reads and writes but few sequential I/Os.

OLAP is part of the broader category of business intelligence, which also encompasses relational databases, report writing, and data mining. A typical OLAP workload is generally characterized by significantly less complex queries, with a larger volume of data, for the purpose of business intelligence or reporting. Whereas OLAP systems are mostly optimized for reading, and the workload is characterized by sequential reads. These workloads are the focus of the thesis.

\begin{figure}[htbp]
	\centering
	\includegraphics[width=1.0\textwidth]{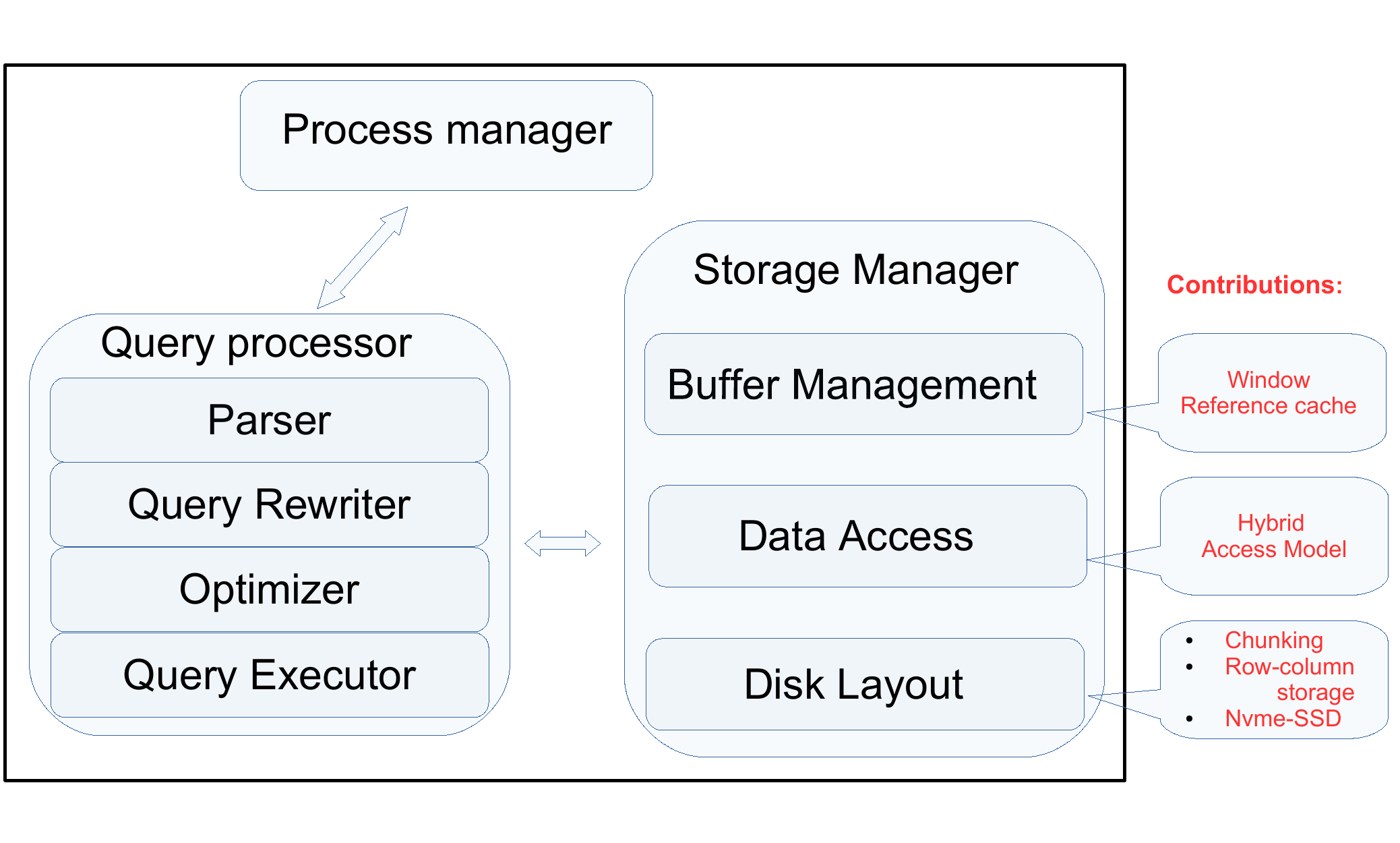}
	\caption{Database Components.\cite{dbreadings,datastage}}
	\label{fig:database}
	
\end{figure}

The most mature database systems in production are relational database management systems (DBMSs), which serve as the backbone infrastructure of many applications. Analytical databases are built for optimizing analytics through data storage and hardware usage. To attain a faster query performance and better maintenance and scalability, analytical databases are optimized by applying column-based data storage, in-memory hardware usage, integrated functions, or other appliances, such as architecture concepts\footnote{https://bi-survey.com/analytical-databases}.

This study focuses on analytical database systems. A typical analytical database system has three main components~\cite{dbreadings}, as illustrated in Figure ~\ref{fig:database}. The process manager is responsible for managing various tasks in the system. The manager communicates with system clients, encapsulates user requests, transmits the requests to the other components, and returns the results to the client. The query processor functions as the central component and consists of a parser, query rewriter, optimizer, and query executor, all of which operate together to translate and transform the user requests into an efficient query plan for execution. The final component is the storage manager, which involves a disk layout, access methods, and buffer management. The storage manager is responsible for long-term data storage. 

Storage often creates a performance bottleneck in the database system, especially for analytical databases systems, which are the cornerstone of many business intelligence architectures. An inefficient storage system often creates a performance bottleneck in the database system, especially for analytical databases systems, which are the cornerstone of many business intelligence architectures.

\section{Storage Manager Overview}
In a database system, data is placed on inexpensive secondary storage (i.e., disks). However, the speed of accessing the data is relatively slow. Reading the same amount of data from a disk is much slower than retrieving it from memory. Moreover, disks have different performance characteristics for different read workloads. For sequential I/Os, a disk has a high transfer bandwidth, but the speed dramatically decreases for random access to the data. To speed up access to the data on disks, pioneers of the database system designed the storage manager. The performance of storage manager is often crucial to the performance of the entire system. Demand for high performing storage managers in analytic database systems continues to increase. The goal of this research is to investigate techniques to improve the performance of the storage manager. A typical storage manager usually includes three main functional layers: buffer management, data access, and disk layout (Figure\ref{fig:database}\cite{thesis:wang}).


\begin{itemize}
	\item \textbf{The Disk Layout Layer} The disk layout layer primarily manages the on-disk data structures. The goal of this layer is to improve I/O efficiency by reducing the mechanical movement of the disk arms. Two choices need to be made. The first choice relates to data organization, and the options are row-store or column-store systems~\cite{monetdb:overview}. The second choice regards the persistent storage interface. This interface can either interact directly with the device drivers for disks~\cite{raid} or use typical operating system (OS) file facilities.

	\item \textbf{Data Access Layer} The data access layer is responsible for triggering data generation and scheduling its workflow. The main objective of this layer is to efficiently coordinate the different components of the storage manager to maximize I/O speed. The method often involves push- or pull-based models~\cite{datapath,volcano}.    
	
	\item \textbf{The Buffer Cache Layer} The buffer cache layer refers to an in-memory buffer called the buffer cache used to store popular disk data. Due to space limitations, a buffer data replacement algorithm regulates the data eviction. Many researchers have focused on the design of the replacement policies. Since the database workload involves using nested loop joins to scan and rescan a heap file larger than the buffer pool, the OS page replacement policy (e.g., Least Recently Used [LRU]) is inefficient~\cite{dbreadings}. Many improved schemas have been proposed, such as the LRU-2~\cite{lru-k}, Cooperative-Scan (CS)~\cite{cooperativescan}. The performance of the buffer cache layer is crucial to that of the entire database system. 
\end{itemize}

The storage manager in most database systems works as a stand-alone component that is shared by different queries. However, the storage manager often has no pieces of knowledge of running queries. To improve performance, the storage manager reads and caches the data at the OS page level. The  Cooperative-Scan~\cite{cooperativescan,abm} presents a novel operation model for the storage manager. All information concerning the data pages is registered in the storage manager prior to the query execution. This mechanism brings query-level metadata to the storage manager to produce a more efficient data access plan. The authors designed both a data structure to store page-level information and a set of algorithms, including estimating processing time, refreshing the priority and eviction policy of the data pages.  

\section{Solid-State Drive}
An SSD is a storage device that uses integrated circuit assemblies to store data persistently. It is also sometimes referred to as a solid-state disk, although SSDs do not have physical disks. SSDs primarily use traditional hard disk drive (HDD) interfaces, such as Serial ATA (SATA) and SAS, greatly simplifying the usage of SSDs in computers.

\begin{figure*}[tbp]
	\begin{center}
		\begin{minipage}[t]{.5\textwidth}
			\includegraphics[width=\textwidth]{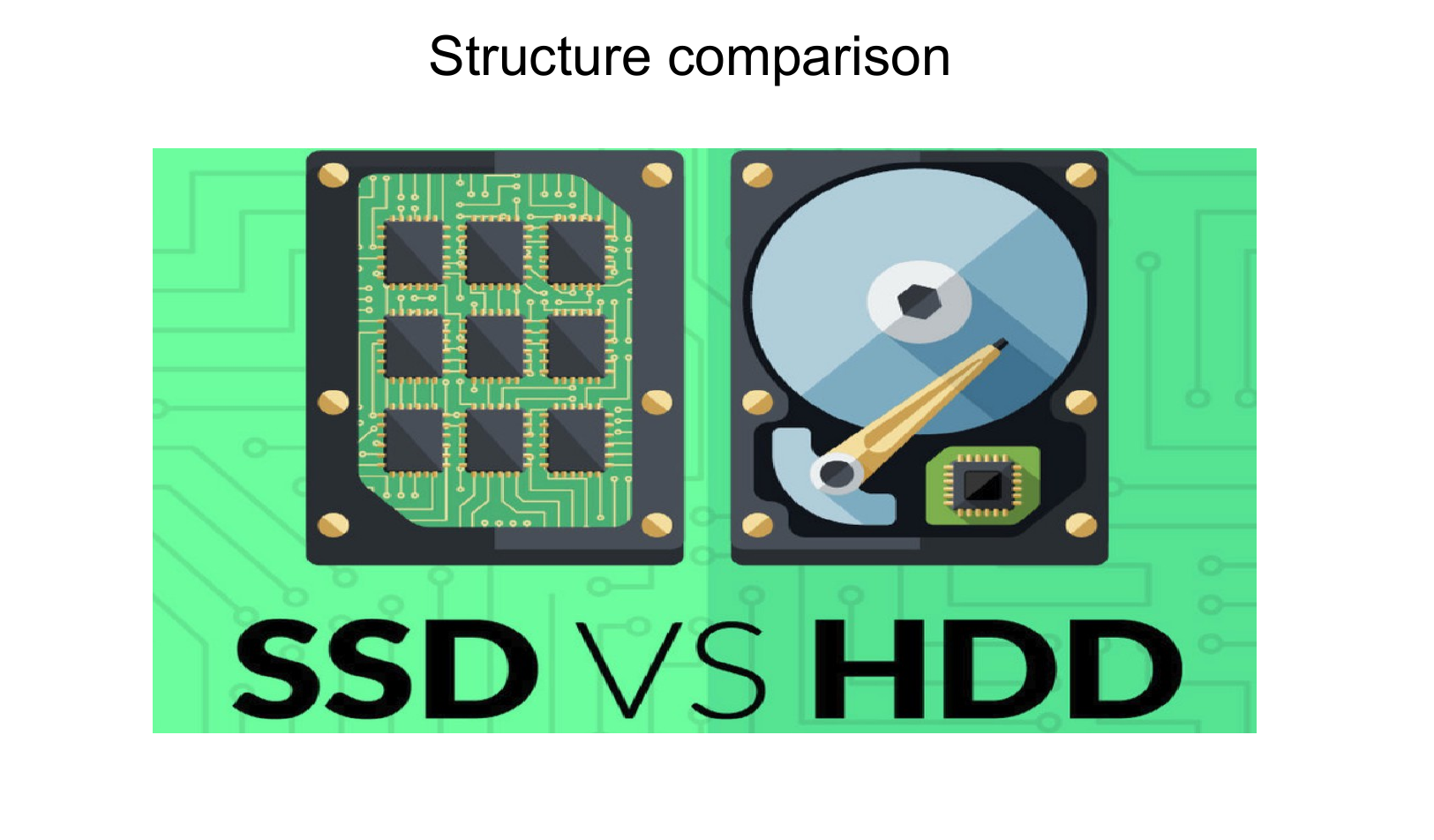}
			\caption{Internal structure comparison.}
			\label{fig:ssdvshdd}
			\footnote{https://tirto.id/membandingkan-kinerja-perangkat-penyimpan-data-ssd-vs-hdd-crSk}
		\end{minipage}\hfill
		\begin{minipage}[t]{.5\textwidth}
			\includegraphics[width=\textwidth]{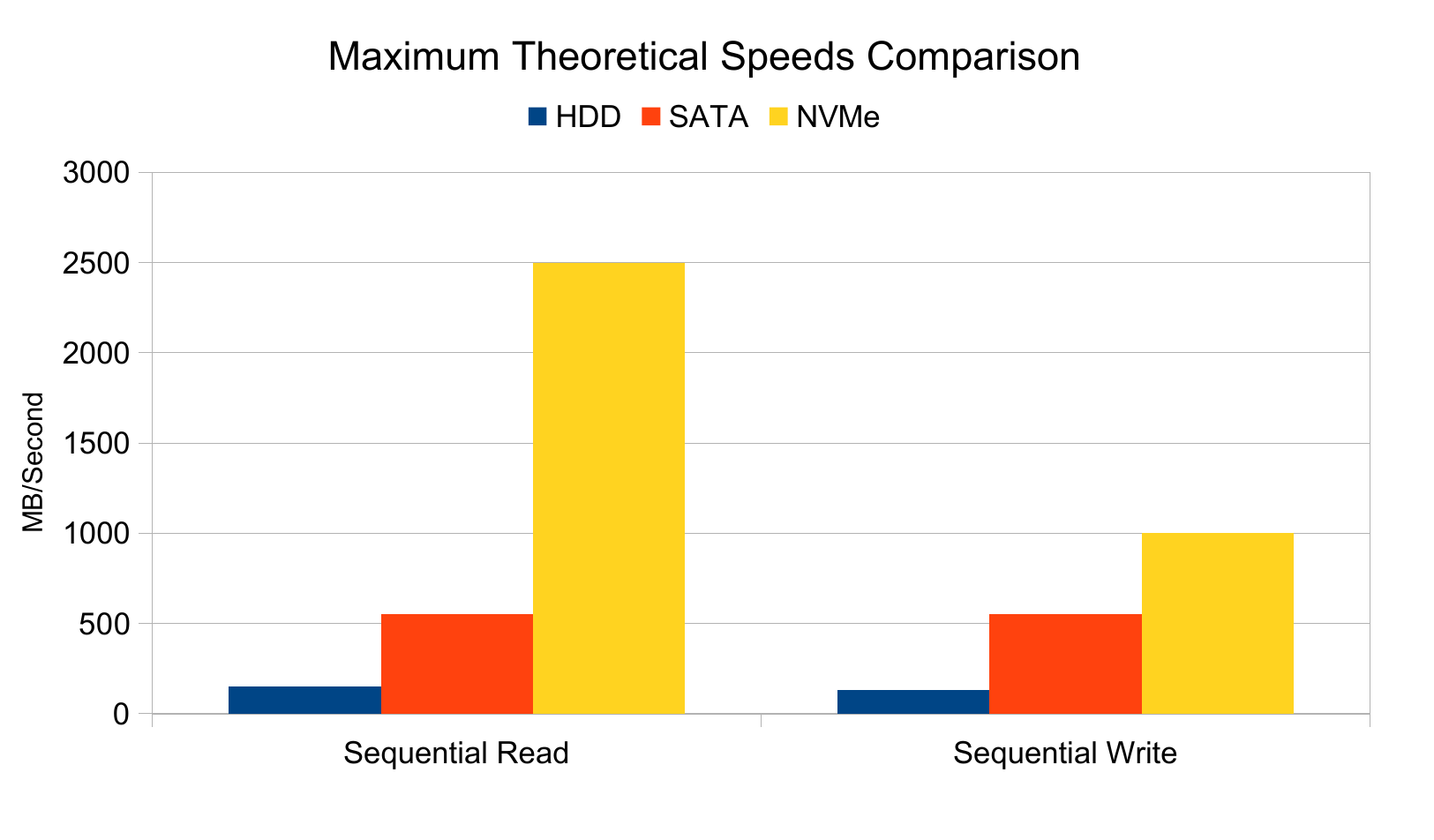}
			\caption{Maximum Speeds of SSD and HDD.}
			\label{fig:nvme}
			\footnote{https://www.atpinc.com/blog/nvme-vs-sata-ssd-pcie-interface}
		\end{minipage}\hfill
	\end{center}
\end{figure*}

Figure~\ref{fig:ssdvshdd} demonstrates the difference in internal structure between SSDs and HDDs. The absence of moving mechanical components distinguishes SSDs from conventional electromechanical drives, such as HDDs or floppy disks, which contain spinning disks and movable read/write heads. Compared with electromechanical drives, SSDs are typically more resistant to physical shock, run silently, and have quicker access times and lower latency.

With the high throughput offered by SSDs, multi-SSD volumes have become an attractive storage solution for big data applications. Flash-based SSDs provide considerable benefits compared to HDDs for such applications due to their high random and sequential performance. It has become increasingly common to deploy multiple SSDs to support a wide variety of applications, such as graph analytics, machine-learning, and key-value stores. 

\textbf{SATA} SATA is a computer bus interface that connects host bus adapters to mass storage devices, such as mechanical HDDs or SSDs. When a computer accesses this type of SSD, it must pass all I/Os through its own SATA controller. Although the speed of the SATA channel increases to 6 Gbps, it is still too low to maximize the SSD performance.

\textbf{NVMe SSD} An NVMe SSD is an interface specification optimized for NAND flash and next-generation solid-state storage technologies. This interface is defined to efficiently support the needs of enterprise and client systems utilizing PCI Express (PCIe) SSDs. 

Compared with other interfaces designed for mechanical storage devices, NVMes reduce latency and deliver a higher number of I/O per second. NVMes also offer performance across multiple cores for quick access to critical data, scalability for current and future performance, and support for standard security protocols.

The figure~\ref{fig:nvme}\cite{nvme} compares a traditional mechanical HDD, an SATA SSD, and an NVMe SSD. The results indicate that while SATA delivers four times the performance of mechanical HDDs, NVMe trumps SATA with five times the sequential read/write performance, enabling the fastest performance and maximum theoretical speeds over any other storage protocol.

\section{Problem Definition}\label{problem:def}
The TPC-H benchmark\footnote{http://www.tpc.org/tpch/}. is a standard method of evaluating database system performance. It contains a series of tables, such as \textbf{ORDER} and \textbf{LINEITEM}, to represent standard industry transactions. As defined by the benchmark specifications, \textbf{LINEITEM} records the transaction details in the table. To clarify, we will consider a representative example, which consists of three queries\ref{fun:queries} in table \textbf{LINEITEM} from the TPC-H benchmark.

\begin{equation}\label{fun:queries}
\begin{aligned}
& {SELECT\ \ SUM(quantity)\ \  FROM\ \  lineitem\ \ WHERE\ tax\ <\ 0.09}\\
& {SELECT\ \ AVG(extendedprice)\ \ FROM\ \ lineitem\ \ WHERE\ tax\ >\ 0.08}\\
& {SELECT\ \ SUM(extendedprice * (1-discount) * (1+tax))\ \ FROM\ \ lineitem}
\end{aligned}
\end{equation}
Here are three queries to calculate some statistic information by the standard aggregation queries such as \textbf{SUM} and \textbf{AVG}, combining with condition filters. We can see clearly that these queries share a partial set of attributes. For example, the tax attribute is accessed by all the queries, and the extended prices are read by Query 2 (Q2) and Query 3 (Q3). This study addresses the problem of \textit{\textbf{offering an efficient data access plan for executing batch queries}}. Our objective is to design a comprehensive solution that provides efficient data access when the workload consists of batch queries.

\begin{figure}[htbp]
	\centering
	\includegraphics[width=1.0\textwidth]{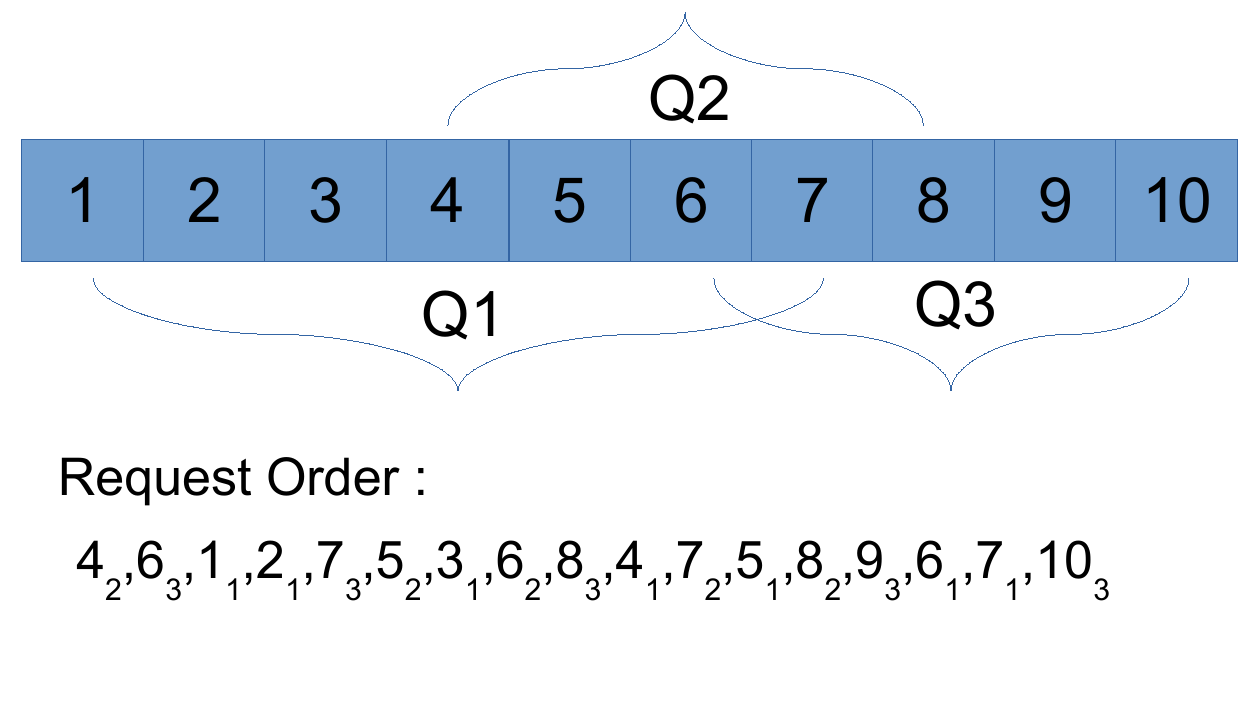}
	\caption{Example of data access plan.}
	\label{fig:pages}
	
\end{figure}

\section{Contributions}
The processing unit in a traditional storage manager is the data page, which is the fixed size of the data, such as 4K. For a given query, whether simple or complex, the storage manager reads a list of pages sequentially and pushes them back to the execution engine for processing. The cache buffer applies the LRU or Least Frequently Used (LFU) for those data pages based on their access orders. Consider the example queries~\ref{fun:queries}, assuming the \textbf{LINEITEM} table contains 10 data pages in figure~\ref{fig:pages}. Due to the conditional filter, query $Q_1$ must read 7 Pages 1 to 7, ; some of these pages are shared by query $Q_2$, which cares Pages 4 to 7. Query $Q_3$ accesses Pages 6 to 10 for processing.

If the queries arrive simultaneously, the storage manager must read the data pages for them. Figure~\ref{fig:pages} describes the page access order for batch queries. The number indicates the data page, and the subscript defines which query made the request. In addition, the cache size is limited; the maximum number of cached pages is three. Under the LRU strategy, the hit rate of the cache management system is zero because it does not consider any query information. There are no repeated numbers in any length-3 windows, so the performance of the storage manager is poor.

\begin{figure}[htbp]
	\centering
	\includegraphics[width=1.0\textwidth]{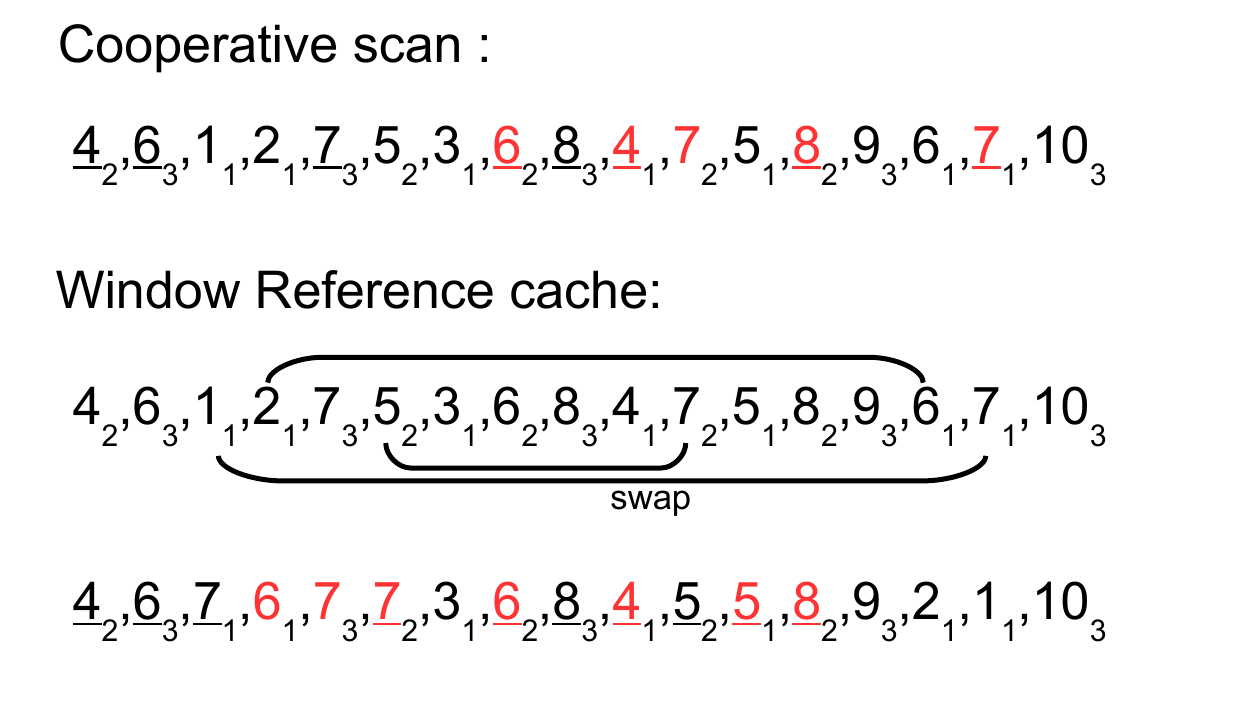}
	\caption{The comparison of cache strategies.}
	\label{fig:swap}
	
\end{figure}

The CS~\cite{cooperativescan}, works as a central controller that aims to improve the performance of buffer management, especially for batch query execution. When queries are registered to the database system at the beginning, the CS utilizes a set of data structures to collect the necessary information about every data page for each query. The CS plans to accurately estimate the consuming time for each data page in batch queries and then assigns those data pages being accessed soonest with a higher priority than those needed later. By this mechanism, the CS can efficiently increase the hit rate of the cache buffer, especially for the batch queries’ engine. In the example from figure~\ref{fig:pages}, the CS performs the same request order more efficiently. Figure~\ref{fig:swap} shows its decision process; the numbers underlined in black indicate that the data is read from a disk and will be cached in the buffer. A red number indicates that the data can directly get from the cache; an underline means this page can be evicted at any time. Pages 1, 2, 3, 9, and 10 are only used once, so those pages will not be cached after reading. The result shows that the CS can increase the cache hitting rate from 0 (LRU) to nearly $30\%$.

Inspired by the CS, the operation model for the storage manager described in this thesis collects the query-level information for data access plan generation. We studied the storage manager in database systems and identified sets of optimizations for the buffer cache, data access, and disk layout layers. This process included analyzing the characteristics of workloads running on the database system at the beginning and identifying performance bottlenecks of the storage manager in different layers.

Although the CS functions well in some scenarios, there are many aspects to improve. One area is the accuracy of estimation; the process time can be reckoned accurately if the system is stable. However, if the workload varies between I/O bound and CPU bound during the process, the variance of processing time for each page could be enormous, resulting in inaccurate estimates. Additionally, the CS utilizes the query-level information for eviction but not for data access plan generation. 

To improve the performance, we preserved the data reference information, such as how many queries will process the data, and proposed a defined request window that contains the data being accessed soon. The priority value of a data unit is decided by two factors: the reference number (how many total queries need this data) and the in-window number (how many are being called soon). To serve the coming queries more accurately and dynamically, the second factor is given a higher weight. We also utilized the query-level information to reorder the data access in the window, which could merge all the data requests of batch queries at the same time. This technique enhanced the data locality and increased the size of the data for each I/O operation, improving I/O performance. The figure~\ref{fig:swap} provides an overview of the primary concepts proposed by this study. We used the query-level information to improve the eviction algorithm, in addition to reordering the data requests inside a single query to maximize data sharing between queries. The results indicate that our proposals could increase the cache hitting rate from $30\%$ to $47\%$. Finally, we used experiments to evaluate the new methods and algorithms; Section 4 describes the results of these experiments.

The main contributions of this project can be summarized as follows:
\begin{itemize}
	\item We performed the research on a state-of-the-art storage manager and made a full understanding of its architecture, implementation, and functionality. The most important aspect was locating bottlenecks under various conditions.
	\item We improved I/O performance through integration with an NVMe SSD, which has a much a higher performance than a mechanical HDD.
	\item We utilized a chunking strategy to partition data and combine the column store to organize the data on a disk. We also ascertained the optimal chunking parameters to improve I/O performance. 
	\item We proposed a high-throughput model (HighTh) to modify the data generation plan dynamically according to the real-time workload. This proposed model combines the advantages of pull- and push-based models to improve performance. 
	\item We designed and implemented an efficient cache management component (i.e., a window reference cache) in the storage manager. This component uses application-level information to prioritize data in the cache for eviction.
	\item We ran extensive experiments to evaluate the performance of the new storage manager. Compared to the CS model, the HighTh storage manager increased the total execution time and reduced the number of I/O read requests sent to the disk.
\end{itemize}

%
%
%
%
%
%
%
%
%
%

\section{Thesis Organization}
The rest of the thesis is organized as follows. Chapter \ref{ch:storageManager} describes the art-of-shell storage management system in details. Chapter \ref{ch:smoptimization} introduces our proposals in different layers to optimize the storage manager. Chapter \ref{ch:experiment} shows the experiment results of performance evaluation of our new storage manager in different configurations. Finally, the conclusion in Chapter \ref{ch:conclusion} summarizes this study and also the future work directions.
\end{large}
\pagebreak

\chapter{Storage Manager}\label{ch:storageManager}
\begin{large}
The \textbf{Storage Manager} is responsible for organizing data on disks and reading and delivering the data to the execution engine for processing. The storage manager operates as an independent component that reads data asynchronously from the disk and pushes it for processing. Magnetic disks remain the dominant medium for secondary storage in database systems. Since the disk access time is much slower than the memory, it is frequently referred to as the bottleneck of performance. Numerous approaches have been proposed to improve the performance of the storage manager. These approaches can be categorized roughly into three layers: \textbf{The Buffer Management Layer}, \textbf{Data Access Layer} and \textbf{The Disk Layout Layer}. The subsequent sections describe these layers in greater detail.

\section{The Disk Layout Layer}
The disk layout layer mainly manages the on-disk data structures. The design objective of the disk layout management is to improve the disk I/O performance by reducing the mechanical movement of the disk arms. The main idea of this component is to improve performance by organizing data in a proper format, according to the requests pattern.

\subsection{Data Organization}
The fundamental issue in the disk layout layer is data organization, which defines how data is stored in the database. According to the data storage layout, DBMSs can be divided into two categories: column-oriented DBMSs and row-store databases. A column-oriented DBMS stores data tables by column rather than by row. Both column and row databases can use traditional database query languages, such as SQL, to load data and perform queries. Both row and column databases can act as the system backbone and serve data for common extract, transform, load (ETL) functions and data visualization tools. However, by storing data in columns rather than rows, the database can more precisely access the data it needs to answer a query rather than scanning and discarding unwanted data in rows. Query performance is increased for certain workloads involving a small number of fields.

\subsubsection{Row-Stores}
Row-oriented database systems store data in a record-oriented format, where the attributes of a record (or tuple) are placed contiguously in storage. When a query needs to access attributes of a tuple, all the tuple attributes will be accessed regardless of how many are relevant to the operation. The advantages of this type of storage become apparent when a large portion of attributes needs to be accessed. With a row-store mechanism, the disk pushes all the fields of a record out to the disk in a single write, which executes a high write performance. Therefore, row-store architecture databases are called write-optimized systems. These types of systems are especially effective in OLTP-style applications. To illustrate the workflow, we will use the stock transaction records as an example.

\begin{figure}[htbp]
	\centering
	\includegraphics[width=0.85\textwidth]{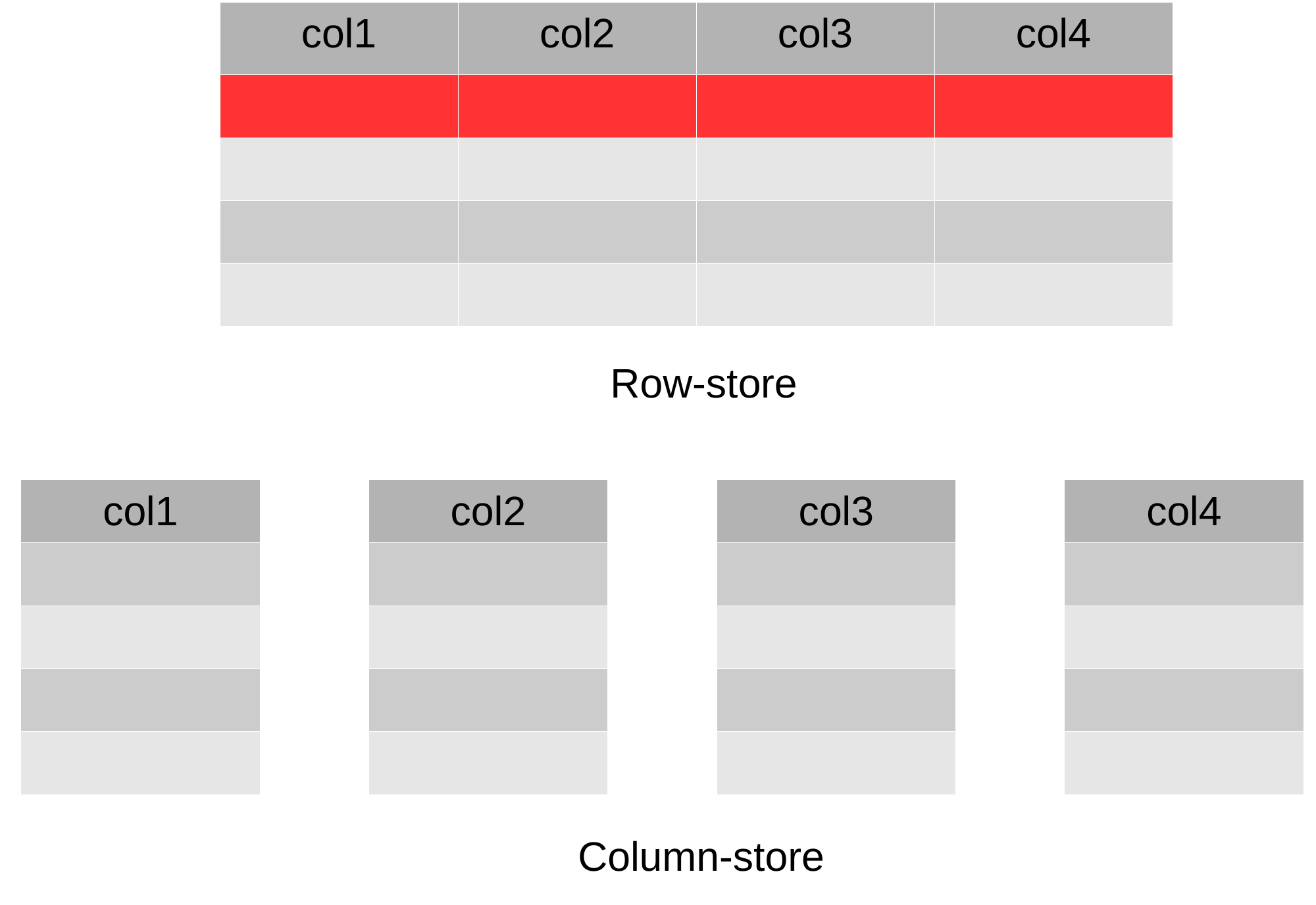}
	\caption{Row-store vs Column-store.}
	\label{fig:row-column}
\end{figure}

In the figure\ref{fig:row-column}, each client has a record with its basic information—name, address, and phone number—in a single table. It is common for each record to have a unique identifier. Often, this identifier is an account ID. Another table stores the stock transactions. Each transaction is uniquely identified by a field transaction ID. Each operation is associated to one account ID, but each account ID is associated with multiple transactions. This one-to-many relationship is a classic example of a transactional database.

When running a query, the system accesses a large amount of data before determining what information is relevant to the specific query. To determine the account\_number, first\_name, last\_name, stock, and purchase\_price for a given time period, the system must access all the information for the two tables, including fields that may not be relevant to the query. The system then performs a join to relate the data from the two tables, after which it can return the information. This process is inefficient at scale in a row-store database.

To alleviate this problem, the row-store DBMS stores along with auxiliary B-tree indexes on attributes in the table. Such indexes can be primary, where the table rows are stored in as close to sorted order on the specified attribute as possible, or secondary, in which case no attempt is made to keep the underlying records in order on the indexed attribute. The index is very efficient, especially for update and range queries. With the help of indexes, the former query could avoid reading unnecessary record tuples and speed up the query process. 

Since row-store databases handle data horizontally (accessing rows of a table resembles reading the table horizontally), they are more suitable for updates and insertions. In row-store systems, the entire row is written in a single operation. This makes row-store systems preferable for OLTP-oriented databases since OLTP workloads tend to be more loaded with interactive transactions, such as retrieving every attribute from a single entity or adding entities to the table. Some of the best-known databases that store data using this technique are Oracle, IBM DB2, Microsoft SQL Server, and MySQL.

\subsubsection{Column-Stores}
The concept of vertically partitioning database tables to improve performance has existed for a long time. MonetDB\cite{monetdb:overview} systems are the pioneers of the modern column-oriented database system.

A column-store database differs from a traditional row-store database in how data is stored in the memory. In a column-store system, attributes depicted by column are stored contiguously (Figure\ref{fig:row-column}). Column-store systems vertically partition the database into a collection of individual columns stored separately. By storing each column separately on disks, these column-based systems enable queries to only read the relevant attributes, rather than having to read entire rows from the disk and discard unneeded attributes once they are in memory. A similar benefit is true when transferring data from the main memory to the CPU registers, improving the overall utilization of the available I/O and memory bandwidth. Taking the column-oriented approach to the extreme allows for numerous innovations in terms of database architectures.

Since data is typically read and written from storage in the form of blocks, a column-oriented approach means that each block that holds data for the transaction table holds data for only one of the columns. In this case, a query that computes, for example, company revenue for the previous year would only need to access the price and date columns, and only the data blocks corresponding to these columns would need to be read from storage.

Whether a column-oriented or a row-oriented layout is a better fit depends on the access patterns in the workload. When a query needs to access only a single record from the disk, a column-store must seek several times to read this single record. In contrast, in a standard row-store database, if a query needs to access a single record, only one seek is needed as the whole record is stored contiguously, and the overhead of reading all the record attributes is negligible compared to the seek time. However, as more records are accessed, the transfer time dominates the seek time, and a column-oriented approach performs better than a row-oriented approach. For this reason, column-store systems are typically used in analytic applications with queries that scan a large fraction of individual tables and compute aggregates or other statistics over them.

\subsubsection{PAX}
PAX refers to a storage layout that vertically partitions the records within each page, storing together the values of each attribute in miniature pages. Figure\ref{fig:chunks} generic structure of the internal data layout of a data page. This type of storage structure vertically partitions columns inside each data page. PAX increases inter-record spatial locality by grouping values of the same attribute that belong to different records with minimal impact on the intra-record spatial locality. In addition, PAX incurs minimal record reconstruction costs because it does not need to perform a join to correlate the attribute values of a particular record.

\begin{figure}[htbp]
	\centering
	\includegraphics[width=0.95\textwidth]{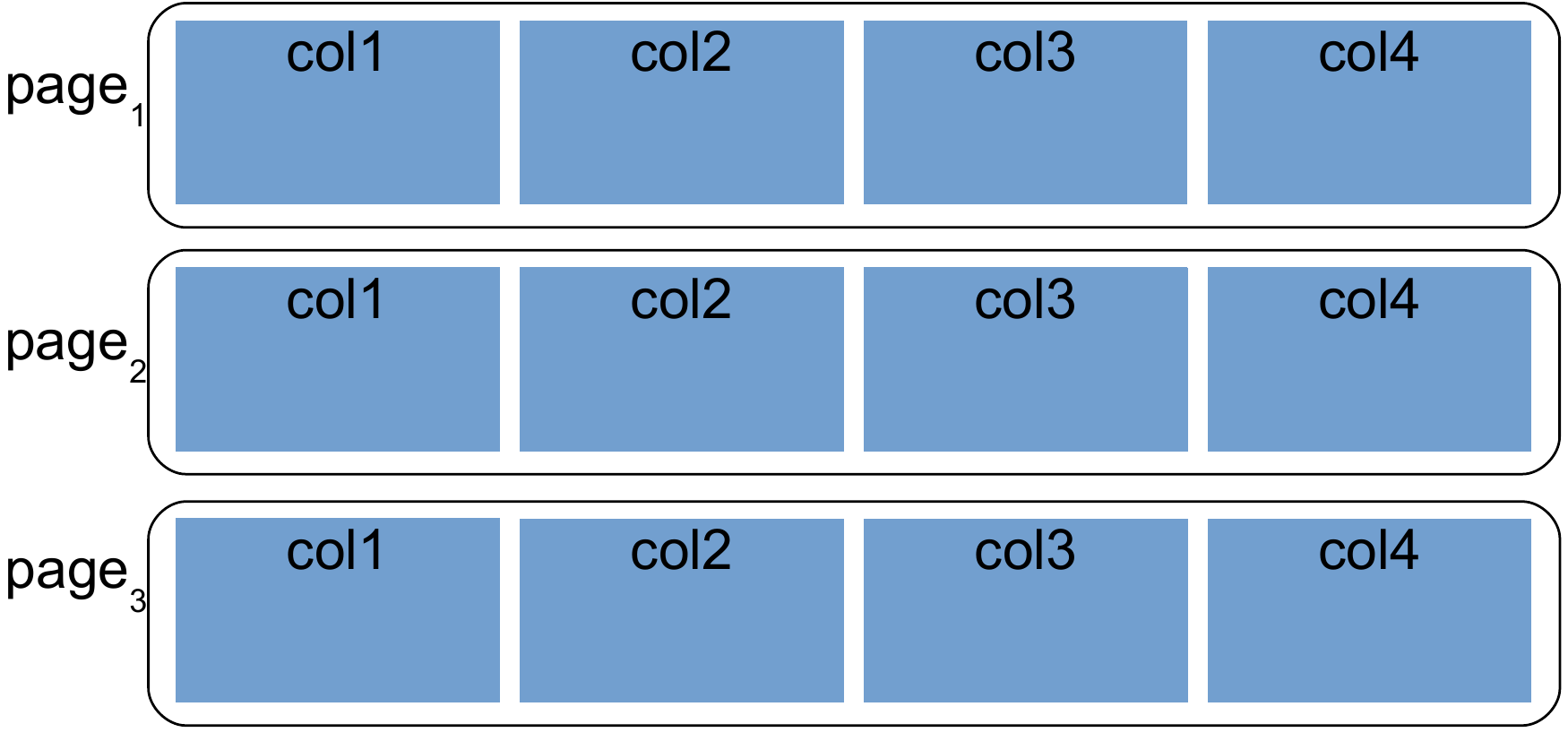}
	\caption{PAX structure.}
	\label{fig:chunks}
\end{figure}

\subsection{File System Optimization}
In addition to data organization, the OS can also play an important role in query performance. The majority of data resides on the magnetic disks for most of the database system. To improve system utilization, the basic read/write unit in the disk typically contains 512 bytes of data. Since every I/O operation on the disk usually involves three mechanical movement costs (seek time, rotational latency, and transfer time), the data access time is slow compared to the memory and CPU. To achieve better performance, the modern file system\cite{operationsystem} contains several elements, including page scheduling, a data cache, and a disk array.

\subsubsection{Page Scheduling}
In database systems, typically many requests are pending concurrently for access to the disk. Since the performance executing sequential disk I/O heavily outweigh the random disk I/O. The page scheduling algorithm aims to reorders the requests at the page level to achieve good I/O throughput. By merging many small requests for random pages into a fewer number of large requests to continues pages, the system can reduce the cost of mechanical movement in I/O operations. However, algorithms generating high I/O throughput often increase the maximum response time by delaying some requests; the tradeoffs between these two factors should be considered.

\subsubsection{Data Cache}
Most disks employ an onboard fast cache to match the speed between the bus and the disk media and cache disk blocks. The disk cache can be used to store prefetched data of the same track to decrease the access time of sequential requests. However, because the requests to the disk are often filtered by upper layer buffer caches, these requests exhibit poor locality in common workloads.

\subsubsection{Disk Array}
Since the performance of a single disk is limited, it is common to organize many disks together to form a RAID\cite{raid}, which has several levels that provide different I/O characteristics and redundancy methods. The most commonly used levels are RAID-0, RAID-1, and RAID-5. In RAID-0, data is distributed across all the disks in the array, providing good performance but no redundancy. In RAID-1, all the data is duplicated on different disks. This level provides good performance, and redundancy is achieved. RAID-5 uses an extra disk-worth of space to store parity information so that a single disk failure does not cause data loss.

\pagebreak

\section{Data Access Layer}
The data access layer defines the workflow of data generation from the disk. The main object of this layer is to efficiently coordinate the different components of the storage manager to maximize I/O performance. Data access is related to read/write data in the database system, which is the most important metric to know to attain optimal database system performance. In traditional database systems, the system must load the user’s data, customarily in the raw format (such as a text format), into the database through the ETL pipeline\cite{ETL}. After loading, the database can process the query on top of this data. Apart from the loaded data, another type of information is stored in the database system referred to as metadata. The metadata describes the loaded data and ordinarily contains its location, size, and structure. The metadata is commonly very small compared to the real data. The loading process transforms the user’s data into a binary format with a specific data type, such as integer, float, and string, which is a compact format.

When the database system receives a query from the client, the storage manager utilizes the metadata to define the position and size of the necessary data using the file system. Next, the data access layer produces a plan to fetch this data in an optimized way and send it to the execution engine for processing.

The data access performance (storage manager) is critical to the efficiency of the entire database system. Therefore, designing an efficient data access layer is important for query optimization. Before offering solutions, the following section describes several basic but important existing solutions for data access to the storage manager.

\subsection{Pull-Based Model}
A database query can be expressed by relational algebra-style query plan languages. Query plans are compositions of operators that can be executed in sequence. However, transferring the computing result between different operators can be expensive, particularly if the intermediate result is large and needs to be stored in memory.

The Volcano Iterator model~\cite{volcano} proposed a solution to this problem for relational database systems. In this model, operators are chained with each other, and tuples are pulled up through operators that are linked by iterators advancing in lockstep. The tuples are produced on demand by latter operators; the intermediate results between the operators are not accumulated.

In the iterator model, the data is lazily generated in each operator by invoking the next method of the source operator by the destination operator. From a different perspective, each operator can be considered as a while loop in which the next function of the source operator is invoked per iteration. The loop is terminated when the next function returns an empty value. In other words, when this empty value is observed, a break statement is executed to terminate the loop execution.

According to the proposed storage manager, in a pull-based model, the scheduler asks for only one chunk at a time. The request for the next chunk is only sent after the first chunk read is complete. All the chunk produce is triggered by the scheduler. The advantage of a pull-based model is that since the data requests are initialized by execution, the produced chunk is necessary and timely, which means that the chunk can be processed as soon as it is produced.

Although pull-based models pipeline the data through pipelining operators, in practice, an issue remains. When the next method of the first operator is invoked, the destination operator should wait until all the operators return the next data.

\begin{figure}[htbp]
	\centering
	\includegraphics[width=0.8\textwidth]{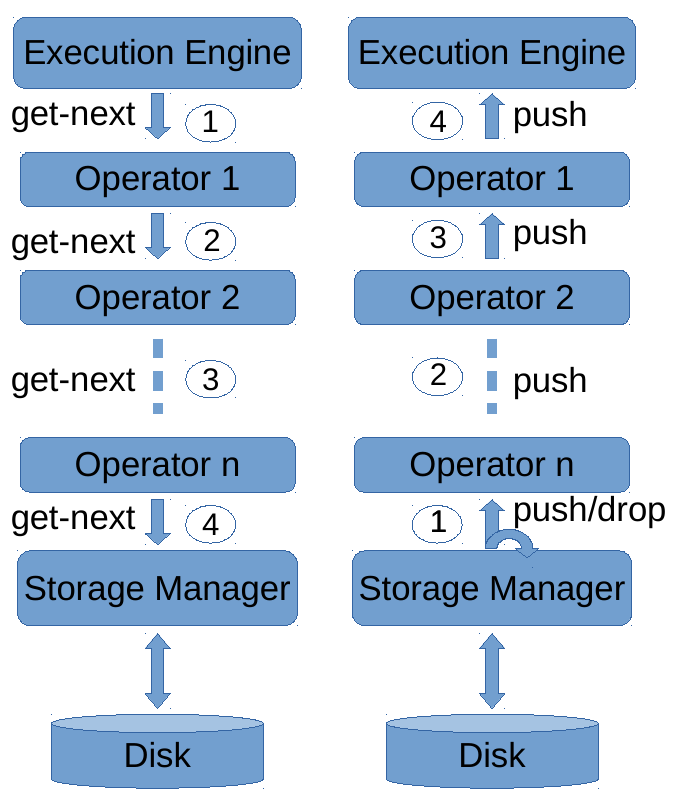}
	\caption{Push vs Pull.}
	\label{fig:pp}
	
\end{figure}

\subsection{Push-Based Model}
The push-based model\cite{pushpull} is widely used in streaming systems. In this model, the control flow is reversed compared to that of pull-based models. The data is pushed from the source operators towards the destination operators without any request from the destination operators. The generated data pushes to the upper level, and once the operator receives the data chunk, it can immediately consume the data and push it to the next operator without waiting. As such, push-based models solve the main problem of pull-based models.

In a push-based model, data is generated at high speed and is pushed to the upper level regardless of whether the produced data is relevant or not. The data passes through each level. When the operator notifies this data is not necessary for processing, the data is dropped. The data will also be dropped, even if it is necessary, if the system or query engine is too busy to process it. The dropped data will be reproduced later by another invoke. Data dropping is a waste of I/O resources.

\pagebreak

\section{The Buffer Management Layer}
In the buffer management layer, data is cached in the buffer cache and managed by a buffer cache management algorithm. The objective of buffer cache management is to reduce the number of disk accesses by maintaining popular data in the buffer cache, thereby reducing the average access time from the disk.

\begin{figure}[htbp]
	\centering
	\includegraphics[width=0.45\textwidth]{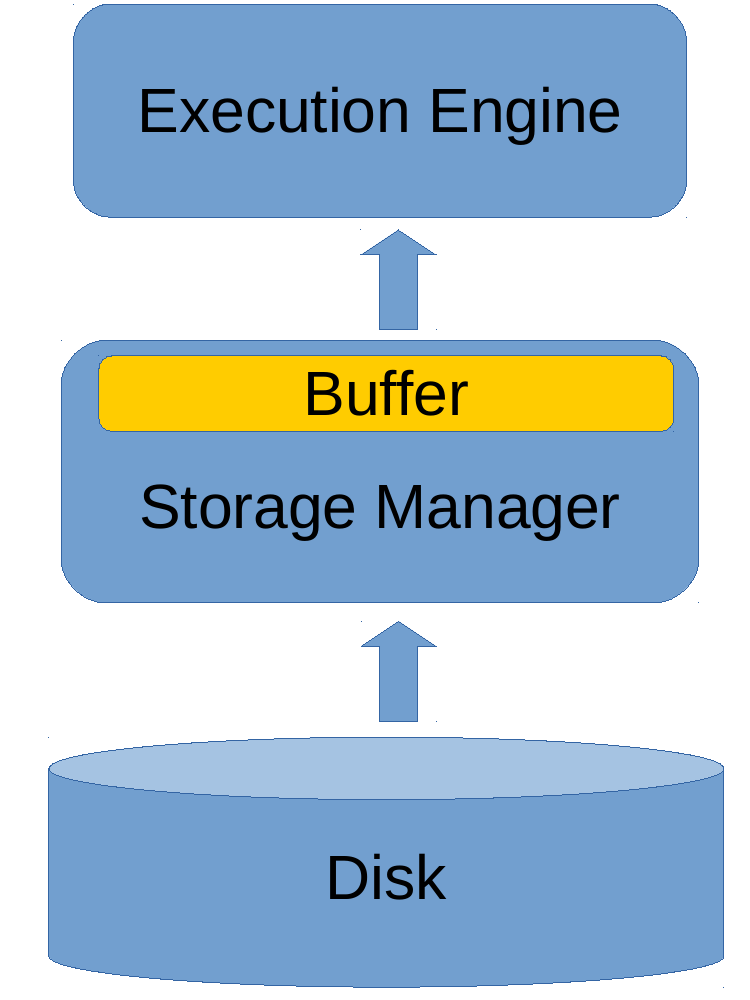}
	\caption{Cache Buffer location.}
	\label{fig:cachelocation}
\end{figure}

\subsection{Data Locality}
The buffer cache replacement algorithm is the most crucial part of buffer cache management. The buffer cache replacement algorithms are designed based on the characteristics of the data access patterns in the buffer cache; the most important attribute is data locality. Understanding the characteristics of the data access pattern is necessary to design an efficient cache replacement algorithm. Usually, there are four types of data locality\cite{datalocality}.

Temporal locality refers to the reuse of specific data and/or resources within a relatively short duration\cite{datalocality}. Temporal locality is called locality-based algorithms, which assume that data visited recently tends to be accessed again in the near future. Temporal locality is observed in many access patterns, such as databases and file buffers.

Spatial locality refers to using data elements within relatively close storage locations\cite{datalocality}. The preferred pattern for this locality assumes that data whose address is nearby to recently accessed data is likely to be visited in the near future. This is the common access pattern in buffer cache management. Spatial locality can be utilized by adopting a large page size.

Sequential locality, a special case of spatial locality, occurs when data elements are arranged and accessed linearly, such as traversing the elements in a one-dimensional array\cite{datalocality}. This is the theoretical basis for prefetching, which can be used to improve performance. 

In looping locality, data is referenced in the same order repeatedly\cite{datalocality}. Contrary to the temporal locality pattern, the most recently accessed data will be accessed again furthest in the future. This pattern could cause locality-based algorithms, such as the LRU, to perform poorly if the loop cannot fit in the buffer cache. Understanding the workload access pattern is the basis for designing an efficient eviction algorithm. To promote an effective algorithm to optimize system performance, a combination of all the data access patterns needs to be considered.

From the above discussion, we can see that understanding the access pattern of workload is the basis for designing an efficient eviction algorithm. To promote an effective algorithm to optimize the system performance, the combination of all the data access patterns needs to be considered.

\subsection{Eviction Algorithms}
Cache eviction policies try to predict which entries are most likely to be used again in the near future, thereby maximizing the hit ratio. The following list describes some of the most common cache eviction policies:
\begin{itemize}
	\item First In First Out (\textbf{FIFO}): The cache evicts the initial block accessed first, regardless of how often or how many times it was accessed before.

	\item •	Last In First Out (\textbf{LIFO}): The cache evicts the most recently block accessed first, regardless of how often or how many times it was accessed before.
	\item Least Recently Used (\textbf{LRU}): The cache evicts the items used least recently first.
	\item Most Recently Used (\textbf{MRU}): The cache evicts the items used most recently first.
	\item Least Frequently Used (\textbf{LFU}): The cache counts how often an item is needed. Items used least often are evicted first.
	\item Random Replacement (\textbf{RR}): The cache randomly selects an item and evicts it to make space when necessary.

\end{itemize}

\subsubsection{LRU}
The LRU policy is perhaps the most popular due to its simplicity, good runtime performance, and adequate hit rate in common workloads. The method the LRU cache employs is simple. A clinet's requests for a resource A is executed as follows: If Resource A exists in the cache, we return immediately. If Resource A does not exist in the cache and the cache has extra storage slots, we fetch Resource A and return to the client, in addition to inserting Resource A into the cache. If the cache is full, we remove the resource that was used least recently and replace it with Resource A.

An LRU cache supports lookup, insert, and delete operations. In order to achieve a rapid lookup, it is necessary to use an implemented hash. Similarly, a linked list enables quick inserts or deletions. To efficiently locate the item used least recently, an order is necessary, such as a queue, stack, or sorted array.

The LRU performs well in common workloads and is the most popular replacement algorithm in real systems such as MySQL~\cite{msql} and PostgreSQL~\cite{postgre}. However, for some special workload access patterns, the LRU performs poorly. For example, when the locality is a looping access but the amount of data in a loop is larger than the size of the total memory buffer, the LRU always replaces the data that will be used soonest. A better replacement algorithm would replace the most recently accessed data since it will be used furthest in the future.

To overcome these problems associated with the LRU, many algorithms have been proposed. Frequently, those algorithms use additional information to make a better eviction decision. The LRU-K ~\cite{lru-k}algorithm, referred to as LRU-2, replaces data whose $kth$ last access has the largest times when the suggested value of K is 2. ARC~\cite{arc} employs a similar data structure and concept but can dynamically adjust itself according to workload changes.

\subsubsection{LFU}
The standard characteristics of the LFU method involve the system tracking the number of instances a block is referenced in memory. When the cache is full and requires more room, the system purges the item with the lowest reference frequency. 

The simplest method to employ an LFU algorithm is to assign a counter to every unit that is loaded into the cache. Each time a reference is made to that unit, the counter increases by one. When the cache reaches capacity and a new unit is waiting to be inserted, the system will search for the unit with the lowest counter and remove it from the cache.

The LFU method appears to be an intuitive approach to memory management, but it has disadvantages. Consider an item in memory that is referenced repeatedly for a short period of time and is not accessed again for an extended period. Due to the rapidness with which it was accessed, its counter increases drastically even though it will not be used again for an extended period. Other units that may be used more frequently are susceptible to purging because they were accessed through a different method.

Moreover, new items in the cache are subject to rapid removal because they begin with a low counter, even though they might subsequently be used frequently. Compared to the LRU, the LFU saves the most commonly used data and picks up long-term patterns. The LFU tends to favor continuing background processes over intermittently used foreground processes. Due to these types of major issues, a pure LFU system is less common; instead, hybrids utilize LFU concepts.

Some modified LFU systems have been proposed to overcome these problems. The Frequency-Based Replacement (FBR)~\cite{fbr} is an improved version of the LFU in which recently accessed data does not accumulate access counts so that correlated accesses are filtered out. The FBR has several parameters that can affect performance but cannot be tuned easily.

\subsubsection{Reference Priority}
In traditional databases, disk scans are often considered inconsequential and therefore use simple LRU or MRU buffering policies. If scans start at different times, these policies achieve only a low amount of buffer reuse. To improve this situation, some systems\cite{qpipe} support the concept of circular scans, which allow queries that start later to attach themselves to already active scans. As a result, the disk bandwidth can be shared between the queries, reducing the number of I/O requests. However, this strategy is not efficient when queries process data at different speeds or a query scans a range instead of a full table.

To overcome these limitations, the CS framework was proposed. This framework transforms the normal buffer manager into an Active Buffer Manager (ABM)~\cite{cooperativescan,abm}. An ABM scans at the start of the execution register their data interest. With this knowledge of all concurrent scans, the ABM can adaptively chose which page to load next and pass to which scan(s) without having to adhere to the physical table order while always trying to keep as many queries busy as possible. To accomplish this, the ABM uses a flexible set of relevance functions that strive to optimize the overall system throughput and average query latency.

A similar approach to concurrent scan buffer management named the Predictive Buffer Management (PBM) has been subsequently proposed. Rather than delivering data out of order as the CS does, the main goal of the PBM is to improve the buffer management policy. The PBM tracks the progress of all scan queries and uses this information to estimate the time of next consumption of each disk page. The PBM employs this knowledge to give pages that are needed soonest a higher temperature so that they will likely be kept in the buffer.

\end{large}
\pagebreak

\chapter{High throughput Storage Manager}\label{ch:smoptimization}
\begin{large}
This chapter proposes several strategies to improve the performance of the storage manager. Section\ref{sec:storageManager} introduces an efficient architecture of the proposed storage manager and describes the core components in detail. Section~\ref{sec:DiskLayout} depicts the strategies to improve the performance of the \textbf{Disk Layout Layer}. In addition, Section~\ref{sec:dataAccess} compares the different data access models of the data access layer and illustrates our proposed methods. Finally, Section~\ref{sec:bufferCache} presents a novel eviction algorithm. To clearly describe our proposed method, we will use the Example\ref{fun:queries}.

\section{Efficient architecture}\label{sec:storageManager}
Figure\ref{fig:storage-arch} illustrates the storage manager skeleton, which applies parallelism to the greatest possible extent. The parallel mechanism is a common optimization strategy that splits a large application into many smaller pieces and allows them to run. This is the most effective way to improve the performance of any complex system. According to the parallelism strategy, the storage manager is divided into four major components: \textbf{TableScanner}, \textbf{ChunkReader}, \textbf{DiskArray} and \textbf{HDThread}. 

These components operate together as an event-driven system that works \textbf{asynchronously}, and they communicate with each other through the message queue. The following paragraphs describe the workflow details, addressing questions related to how the components are able to work together efficiently.

\begin{figure}[htbp]
	\centering
	\includegraphics[width=0.87\textwidth]{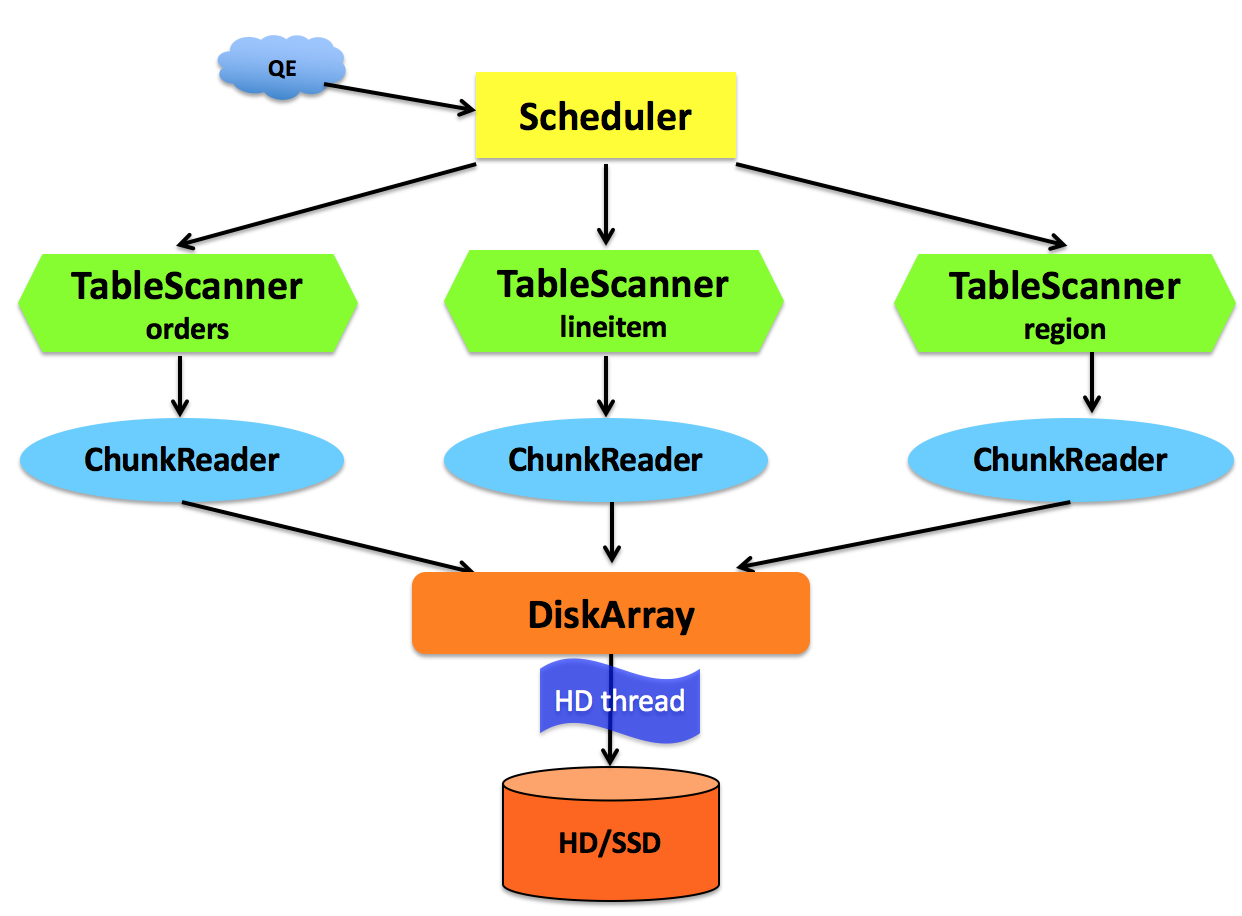}
	\caption{Architecture of Storage Manager.}
	\label{fig:storage-arch}
\end{figure}

Ordinarily, there are two general types of parallelization~\cite{dewitt-paralleldb}: data parallelism, task parallelism is applied in the common practical system.

\subsection{Data Parallelism}\label{subsec:parallelism:data}
Data parallelism is a form of parallelization across multiple processors or cores in parallel computing environments. Data parallelism focuses on distributing data across different processors or cores and emphasizes the distributed nature of the data as opposed to processing. Due to the chunking mechanism, data is assigned across multiple computation nodes in the distributed system or to different disks in the multi-threads system. During the query process, the chunks can be passed to the execution engine immediately after they are produced by the storage manager.

\begin{figure}[htbp]
	\centering
	\includegraphics[width=0.85\textwidth]{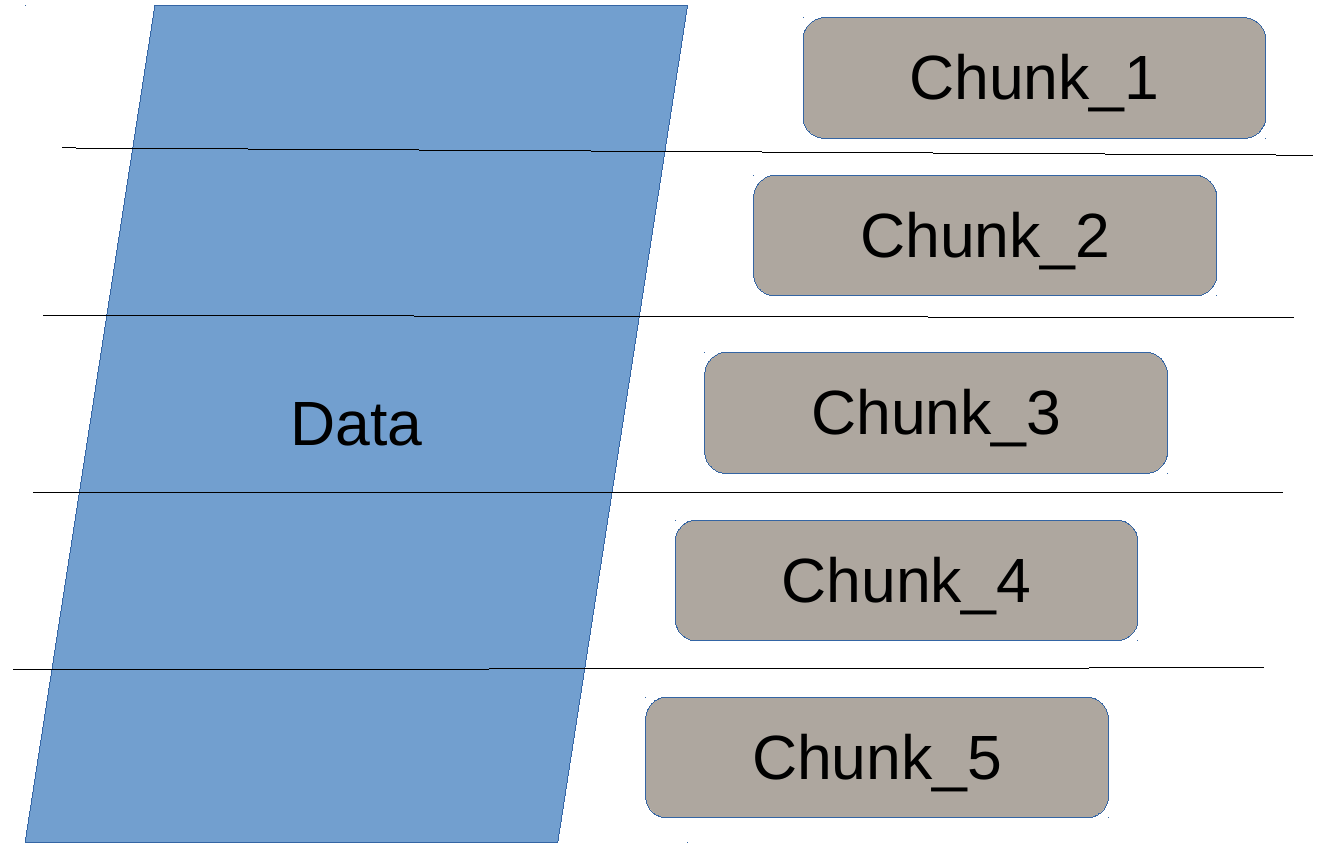}
	\caption{Horizontal partition.}
	\label{fig:filePartition}
\end{figure}
Chunking\ref{fig:filePartition} is an efficient data parallelism strategy. When executing a query, partitions are independently assigned to different execution entities for processing. Since each processing entity works on a considerably smaller dataset, a speed increase proportional to the number of processing workers can be obtained in optimal conditions. Data partitioning can be applied both to the files and to the internal processing representation. We applied the data partitioning strategy~\cite{dewitt-paralleldb} described in by breaking the file into multiple segments of a fixed size in the order of tens to hundreds of megabytes. This strategy has the potential to increase the length of sequential scans and reduce the number of disk seeks. The segment, or chunk, is both the read/write and processing unit.

\begin{figure}[htbp]
	\centering
	\includegraphics[width=0.95\textwidth]{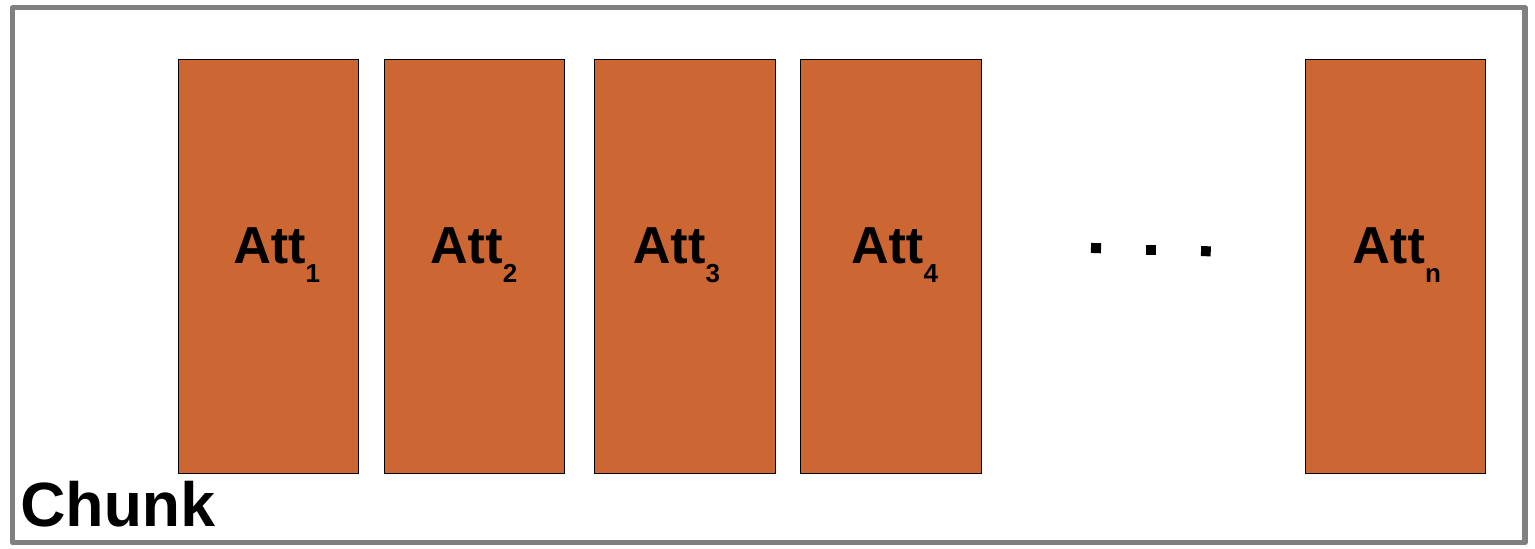}
	\caption{Chunk structure for internal processing.}
	\label{fig:chunk-structure}
\end{figure}

Figure~\ref{fig:chunk-structure} depicts the generic structure of an internal chunk. Inside the chunks, we applied a column-based storage. This type of storage vertically partitions columns inside the chunks, associated with an array of pointers to all the columns. The actual data is vertically partitioned inside a chunk, with each column stored in a separate set of disk blocks. This design can improve the performance for accessing selective attributes and allows only the required columns to be read for each query, thus minimizing the size of reading data. Chunks also serve as the basic unit of parallelism, since different chunks can be sent to different CPU cores for processing.

\subsection{Task Parallelism}\label{subsec:parallelism:task}
Task parallelism\footnote{\url{http://en.wikipedia.org/wiki/Task\_parallelism}} also known as function parallelism or control parallelism, is a form of parallelizing computer code across multiple processors in parallel computing environments. Task parallelism focuses on distributing execution processes, or threads, across different parallel computing nodes. Task parallelism can be applied to improve the performance of the storage manager by assigning the stages identified in Figure~\ref{fig:storage-arch} --\texttt{TableScanner}, \texttt{ChunkReader}, \texttt{DiskArray} and \texttt{HD Thread} --to separate processes or threads. In modern multi-core CPUs, different stages, as well as multiple instances of the same stage, can be executed concurrently. Query processing can be viewed as another task in the parallel task assignment process.

The main functionality of the \textbf{TableScanner} is to generate a data chunk from a table. In the storage manager, multiple table scanner instances run simultaneously. Each scanner works for a specific table, such as the table scanners for orders, lineitem, and region illustrated in Figure \ref{fig:storage-arch}. The \textbf{TableScanner} works at the top level compared to the other components in the storage manager and communicates directly with the \textbf{Scheduler}, which forwards data requests for a query. Then, the tablescanner looks up the metadata, such as the index, and determines the necessary data, which will be transformed into the correct address format—chunk and column IDs—and sent to a lower level component.

The \textbf{ChunkReader} is responsible for constructing the data chunk in the memory. Upon receiving the request from the table scanner, the \textbf{ChunkReader} calculates the size of the requested data, applies for enough space in memory, and transforms the request into another type of address (a list of data page IDs) for a lower level component. After the data is read from the lower level component, the chunk reader compiles those data pages together and sends them as a data chunk to the upper component, the tablescanner.

The \textbf{Disk Array} functions as a dispatcher and result assembler between the \texttt{ChunkReader} and \texttt{HD Thread}. The disk array splits the data read/write request into multiple tiny operations and evenly distributes them to the available disks in a round-robin pattern. The HD thread controls the disk operations. It responds to read/write data to and from the disks in the data pages (4 KB per page).

\section{Enhancement in Disk Layout Layer}\label{sec:DiskLayout}
The disk layout layer is fundamental to the storage management system. It resides at the bottom level of the system, which is the basic component. The proposals in this layer aim to achieve the following objectives:
\begin{itemize}
	\item Increase the I/O bandwidth of the system
	\item Determine the data unit to be transferred between different parts
	\item Organize the data in the disk efficiently
	\item Ascertain a method to efficiently read a data chunk from the disk.
\end{itemize} 

\subsection{New Hardware Deployment}

The simple and efficient way to increase the system I/O performance is to improve the throughput of the hardware system. NVMe is a new protocol that addresses many of the shortcomings of SAS and SATA, including a more direct connection to the processor, optimized I/O channels, and a simplified software stack. NVMe devices are designed to connect to the PCIe root complex, putting them closer to the processor, and residing on what used to be called the Northbridge. This setup reduces latency and raises new ways to communicate with remotely connected devices. NVMe introduces parallelism using multiple I/O queues and greater queue depths—up to 64,000 in both cases.\footnote{\url{https://searchstorage.techtarget.com/feature/NVMe-SSDs-Is-there-a-need-for-all-this-speed}}

Simplifying the I/O stack includes using new signaling methods to indicate when I/O requests are ready to process. This includes the doorbell concept, where the NVMe device signals an I/O completion to the host rather than the host having to continually check the status. This process saves CPU overhead on the server and reduces the time for the software to process each I/O. In comparison, a $7200$ RPM SATA drive manages around $100$ MBps, depending on the age, condition, and level of fragmentation. NVMe drives, in contrast, provide write speeds as high as $3500$ MBps, which is more than seven times greater than the SATA SSDs.

\subsection{Data Organization}

Data organization is related to the layout of data inside the storage, such as on disks. This section introduces the chunking mechanism that partitions data to optimize parallelization. To illustrate the entire workflow, we will reconsider Example\ref{fun:queries} from Section\ref{problem:def}. The $Lineitem$ table contains all the transaction records and has $16$ columns with different data types, such as integer, float, date, and string. With different scaling factors, the size of this file can vary from $1MB$ to more than  $1TB$. We chose a scaling factor of 1.0; the file size was around $1GB$ and the number of tuples was around $6,000,000$. The arrangement for storing the data follows a specific order:

\begin{itemize}
	\item Define the chunk size and partition the file into multiple chunks.
	\item Inside each chunk split the data by columns, which means data from one column of a chunk will be stored contiguously on the disk.
	\item Determine the optimal page size for the storage manager. This size represents the unit for every read/write I/O operation.
	\item For each data unit (chunked, column), store the data in pages continuously on the disk.
\end{itemize} 

\begin{figure}[htbp]
	\centering
	\includegraphics[width=0.95\textwidth]{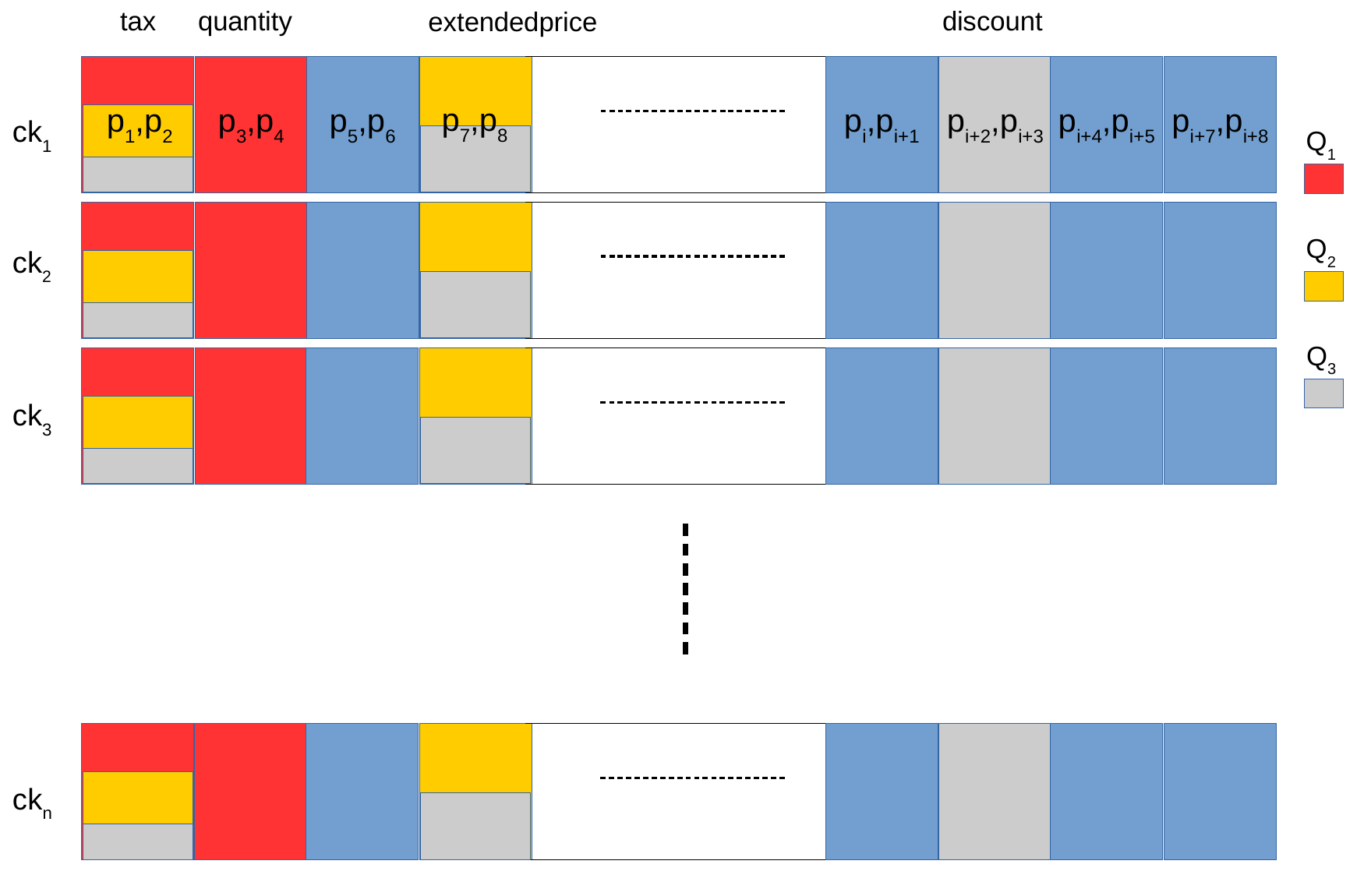}
	\caption{Lineitem Data Organization.}
	\label{fig:lineitem}
\end{figure}

The figure~\ref{fig:lineitem} shows the data organization of the $Lineitem$ table in the proposed system. The table is horizontally partitioned into multiple chunks from $1$ to $n$. Inside each chunk, the columns are vertically split but continuously stored. For example, in Chunk 1, the \textit{tax} column is stored in Pages 1 and 2, Pages 3 and 4 contain the quantity column, and so on. 

There are three types of data structure in this scenario: page, column, and chunk. The minimal unit of I/O operation in the storage manager is a page, but it is different from the pages in the file system. The system page size is often around 4 KB, which is determined by the instruction set architecture, processor type, and operating (addressing) mode. However, the page in the storage manager has more flexibility and generally depends on the query workload. When transferring from a rotating disk, a large portion of the delay is caused by the seek time (the time required to correctly position the read/write heads above the disk platters). Consequently, large sequential transfers are more efficient than several smaller transfers. Transferring the same amount of data from the disk to memory often requires less time with larger pages than with smaller pages. Based on the above description, the page size in the storage manager is much larger than the file system, which is between $2^{22}$ to $2^{24}$  bytes. In the experiment section, we demonstrate that this is the optimal configuration.

The chunk size is another parameter that plays an important role in the data partition process and could heavily impact system performance. Small chunks lead to a more even distribution of data at the expense of more frequent migrations. This incurs a high cost for data routing. Large chunks lead to fewer migrations. This is more efficient both from the networking perspective and in terms of internal overhead at the query routing layer. However, these efficiencies could potentially result in an uneven distribution of data. For a large data set, the data is first divided into multiple data chunks. Assume the total number of tuples is $N$ and the size of the data chunk is $n$, then the number of chunks would be $ceil(N/n)$ with a chunk ID from $0\ to\ ceil(N/n) - 1$. The number of columns is decided by the number of attributions in the data. In this layout, the combination of chunk and column IDs can be used to define the basic unit of data in the system, which is a single column inside a chunk.

According to the description in Section\ref{sec:storageManager}, the storage manager contains four major components, the \texttt{TableScanner}, \texttt{ChunkReader}, \texttt{DiskArray}, and \texttt{HD Thread}. To run a query, these components must cooperate with each other by communicating through messages. The message content among the different components is distinct. For example, to generate a data chunk for $Q_{2}$ (from Example~\ref{fun:queries}), due to the column layout, the storage manager could only read the \textit{extendedprice}, \textit{discount} and \textit{tax}; however, the storage manager has to produce all the chunks because there is no filter in the query. According to Figure\ref{fig:lineitem}, to generate $chunk_{1}$ for $Q_{2}$, the HD thread reads Page 1,2 for column \textit{tax} and Page 7,8 for column \textit{extendedprice}. The disk array then fills the data pages for each column and returns the columns to the chunk reader, which constructs the columns into a data chunk. When the table scanner receives intel that the data chunk has been read back from the chunk reader, the table scanner pushes the chunk to the scheduler for computation.

\section{Strategies in Data Access Layer}\label{sec:dataAccess}

The goal of the data access layer is to efficiently produce data from the lower disk layout layer for multiple queries. Since the processing unit in the proposed system is a chunk, this section describes how data chunks are generated for multiple queries. To produce data for a query, the storage manager determines what data needs to be read with the help of metadata. The data is represented as a list of pair $pair<ChunkId, ColumnId>$. In Example~\ref{fun:queries}, all the queries involve the \textit{tax} column, and both $Q_{2}$ and $Q_{3}$ are interested in the \textit{extendedprice} column. $Q_{1}$ and $Q_{2}$ have different columns to read. Since $Q_{1}$ and $Q_{3}$ have filters for the \textit{tax} column, they may not need to read all the chunks. Each query has its own data list, which might overlap with others. In Figure\ref{fig:lineitem}, the $Q_{1}, Q_{2}$ and $Q_{3}$ data lists are represented in red, orange, and gray, respectively. Areas with multiple colors indicate that the data is needed by more than one query; for instance, the column \textit{tax} data chunks are involved by all three queries.

In our storage manager, we designed an efficient hybrid model to trigger data generation. To illustrate the benefits of this model, we will compare push- and pull-based models using the same examples. As described in Section~\ref{sec:dataAccess}, there are two types of data access models, push- and pull-based. For the pull-based model, data access is driven by the upper level component. The data chunks are generated only when the query engine requests them. The disadvantage of the pull-based model is the high waiting cost for the execution engine. Since $Q_{2}$ and $Q_{3}$ access the same columns (\textit{tax} and \textit{extendedprice}), the storage manager generates the same data chunk twice. However, in the push-based model, data is purely generated by the I/O component. In this model, data usually can be shared by multiple queries. The drawback is that since the push-based model has no information of the upper level component, it is difficult to control the generated data at an optimal speed, especially when there is no computation resource to process the data. All the produced data has to be dropped immediately and read in the future, which wastes I/O operations. The left side of the Figure~\ref{fig:methods-all} illustrates the workflow of the two standard models in the storage manager. The arrow orientation indicates the force of data generation.

\begin{figure*}[htbp]
	\begin{center}
		\subfloat[]{\includegraphics[width=0.5\textwidth]{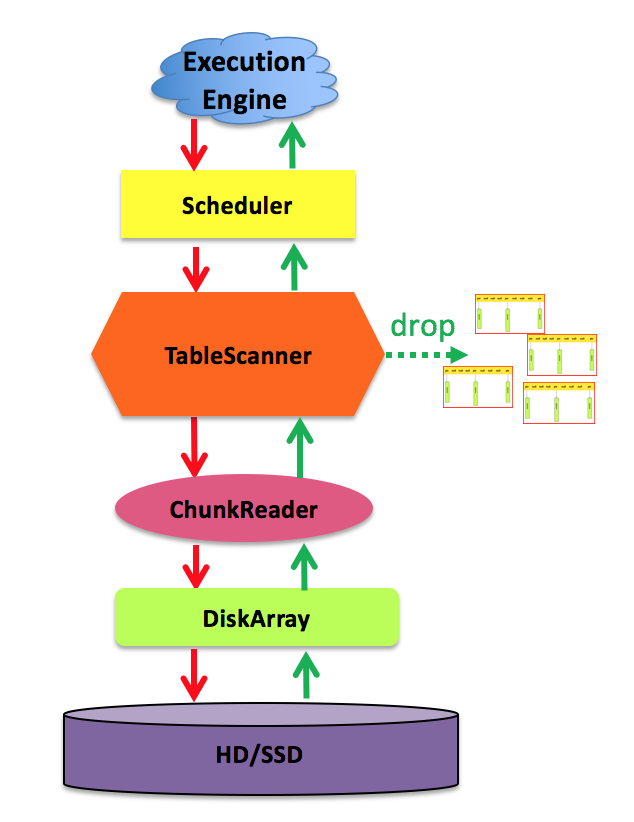}\label{fig:pull}}
		\subfloat[]{\includegraphics[width=0.44\textwidth]{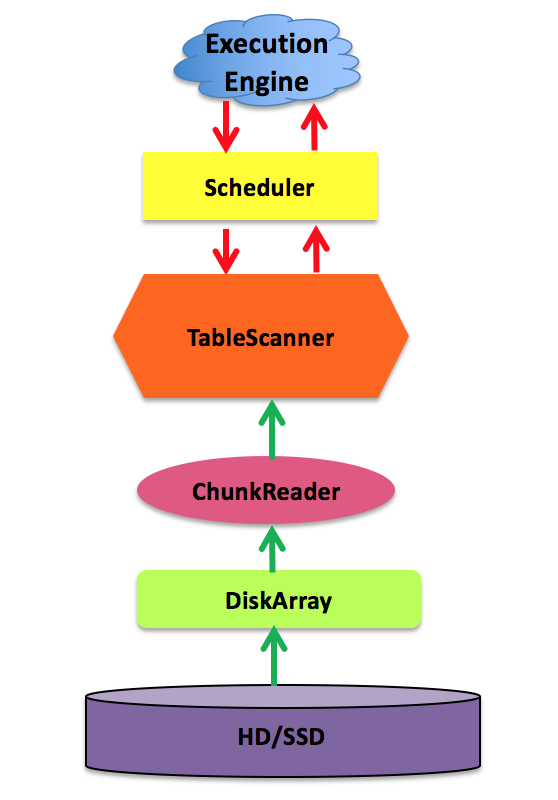}\label{fig:push}}
		\caption{Pull and Push Model (a), Hybrid Model (b).}
		\label{fig:methods-all}
	\end{center}
\end{figure*}
 
To overcome the disadvantages of the pull- and push-based models, we proposed a hybrid model that combines them. The right side of Figure~\ref{fig:methods-all} depicts the hybrid model’s workflow. In the data generation pipeline, the table scanner is the joint component that merges the two models. The upper level components (execution engine, scheduler, and table scanner) initialize the data requests for each query. The data is routinely defined as high-level address, which is a list of $pair<ChunkId, ColumnId>$. The table scanner works intelligently to gathers requests without sending them downstream immediately. Instead, it reorders the requests to make the requested data align with other queries.\textbf{ Data with the same chunk ID (even for different queries that might request different columns) is merged together, indicating that the data will be read through a single I/O operation and processed by multiple queries’ computations}. The reason for this is that columns belonging to the same chunk are stored nearby.

\begin{algorithm}[htbp]
	\caption{Insert\_Request\_List}\label{alg:insertRList}
	\begin{algorithmic}[1]
		
		\REQUIRE Request Data Chunk $chunk$; Request List $RList$
		
		\STATE\label{A3:l1} $chunkId = RList.Find(chunk.id)$;
		\IF{($chunkId > 0$)} 
		\STATE\label{A3:l2} $RList.get(chunk.id).merge(chunk)$
		\ELSE
		\STATE\label{A3:l3} $RList.append(chunk)$
		\ENDIF 
		
	\end{algorithmic}
\end{algorithm}
\begin{figure}[htbp]
	\centering
	\includegraphics[width=0.88\textwidth]{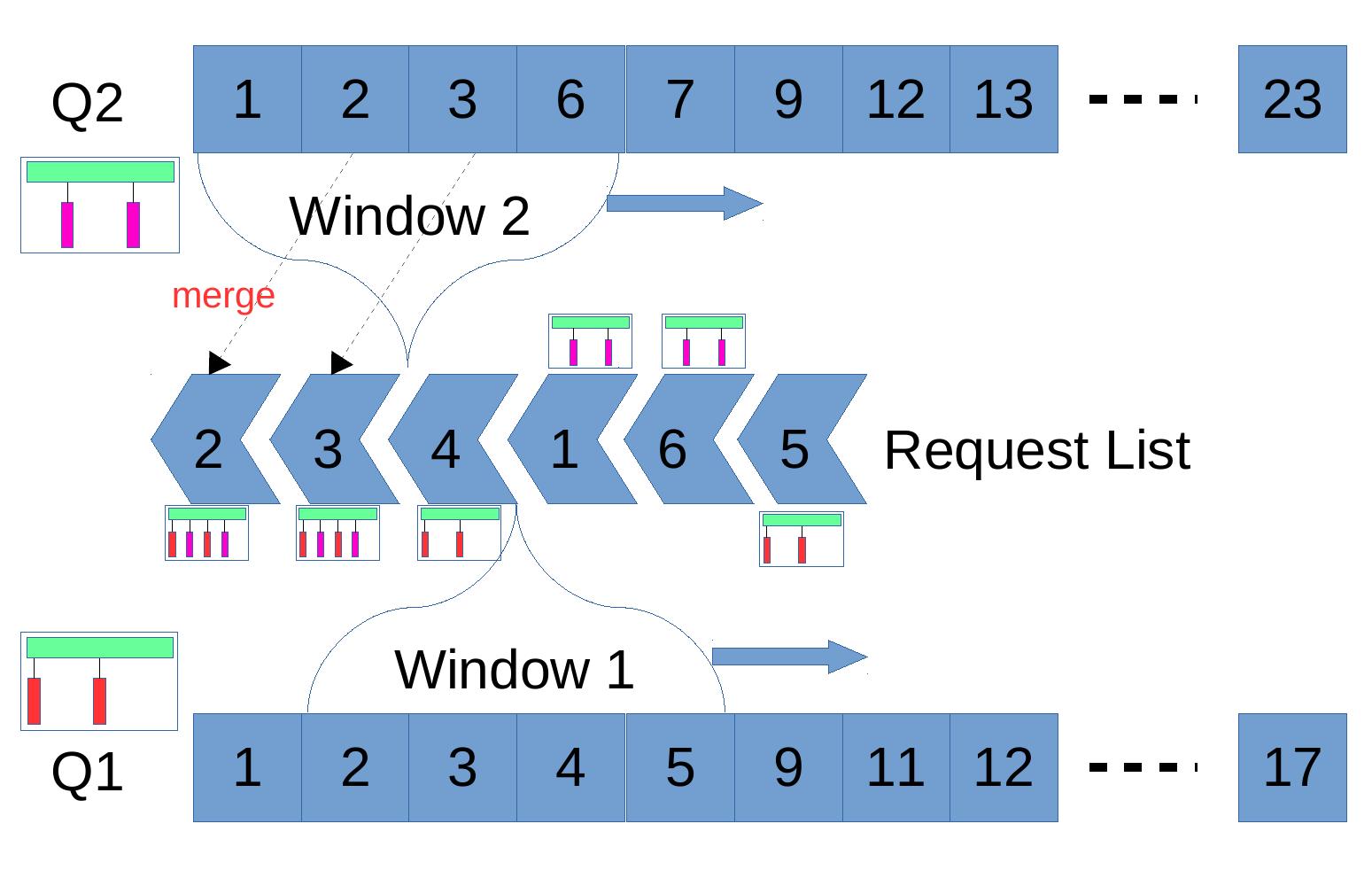}
	\caption{Query Windows.}
	\label{fig:queries}
\end{figure}

Due to the different query process speeds, we cannot modify the request order of the entire data chunk list for all queries at one time. In our example, if the storage manager always generates data chunks with four columns (tax, quantity, and discount), then the I/O performance is efficiently maximizing the I/O sharing. However, if the query speeds are different, it is possible that the slowest query cannot consume the data because of the computation limitation and must drop it. To solve this problem, we designed a request window for each query, such as Window 1 for $Q_{1}$, which moves from beginning of the read list to the end along with the query execution (Figure\ref{fig:queries}). The entire chunk in the window is pushed into the request list, such as Chunk 1,2,3,6 from $Q_{1}$ and Chunk 2,3,4,5 from $Q_{2}$. Additionally, chunks with the same chunk ID are merged; therefore, Chunks 2 and 3 in the request list contain the columns both from $Q_{1}$ and $Q_{2}$. Algorithm~\ref{alg:insertRList} describes the details of merging the requests to the same chunk to maximize the amount of data per chunk read and only append the chunk request if it cannot be merged to an existing chunk. 

Once the chunks inside the request list are optimized, they are continuously sent to lower level components for reading and guarantees that when they are ready, there are sufficient computation resources to process them. The parts below the table scanner include the chunk reader, disk array and HD thread. These components work together in the push-based model; once the requests are defined, the data is continuously produced and pushed to the upper level.

\section{Innovation in the Buffer Cache Layer}\label{sec:bufferCache}
The following subsections describe this study’s solutions to several problems:
\begin{itemize}
	\item Determining the location for the cache component in the storage manager.
	\item Establishing the structure of a basic data unit.
	\item Designing an efficient eviction policy.
\end{itemize}

\begin{figure}[htbp]
	\centering
	\includegraphics[width=0.88\textwidth]{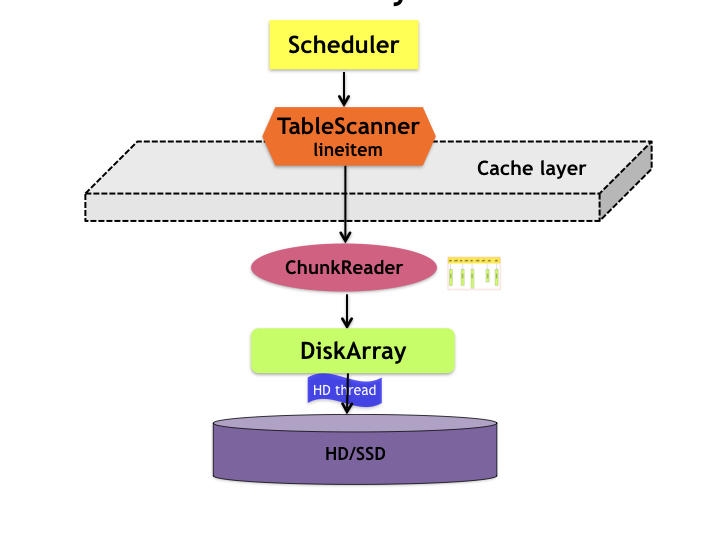}
	\caption{Implement place of Cache.}
	\label{fig:cache}
\end{figure}

The cache system is a widely adopted technique in almost every application today. In addition, it applies to every layer of the technology stack. For instance, a network area cache is used in DNS lookups, and a web server cache is used for frequent requests. In short, a cache system stores commonly used resources (sometimes in memory), and the next time the same resource is requested, the system can return it immediately. This approach increases system efficiency by consuming more storage space. 

To provide efficient access to the database pages, every DBMS implements a large shared buffer pool in its own memory space. The buffer pool is organized as an array of frames, where each frame is a region of memory the size of a database’s basic unit. Units are copied into the buffer pool from the disk without changing format, manipulated in memory in this native format, and later written back. This translation-free approach avoids CPU bottlenecks in marshaling and unmarshalling data to and from the disk. The fixed-sized frames avoid complex memory management of external fragmentation and compaction. Traditionally, the buffer pool has been statically allocated to a set value, but most commercial DBMSs now dynamically adjust the buffer pool size based on system needs and available resources.

\subsection{Location}
According to the discussion in section\ref{sec:storageManager}, the storage manager contains four components  \texttt{TableScanner}, \texttt{ChunkReader}, \texttt{DiskArray} and \texttt{HD Thread}. The cache can be applied between any of those stages. But the cache location will decide the structure of a basic data unit. For example, if the cache was located between \texttt{DiskArray} and \texttt{HD Thread}, then the data in the cache should be disk page, then when DiskArray plan to request data, it will try to fetch it directly from the cache at beginning and then only send the requests to HD Threads when it fails to get the data directly from the cache. In contrast, placing the cache between \texttt{TableScanner} and \texttt{ChunkReader} is much different. Since the table scanner read/write the whole chunk, therefore in its point of view the basic data unit is the data chunk, the table scanner would prefer to get the data directly from the cache, to avoid read it from ChunkReader.  

Although the location of the data cache is flexible, to maximize the unnecessary computation workload, we designed the cache to connect the table scanner and chunk reader, as depicted in Figure\ref{fig:cache}. From the figure we can see that when the scheduler requests the next data for processing, the table scanner initially attempts to obtain the data from the cache. If the data chunk is not in the cache, the table scanner sends the request to the chunk reader to fetch the data from the disk. If the data chunk could be directly read from the cache, it would avoid transforming the request from chunks to columns and disk pages, which involves following multiple steps: mapping the chunk to a calculated memory size between the chunk reader and disk arrays, splitting the memory size into multiple disk pages, reading this data from disks, and assembling it in the opposite direction to compose the data chunk.

A cache is similar to a short-term memory that has a limited amount of space. The system cannot keep all the data in the memory forever. Consequently, we designed an efficient cache eviction algorithm for the system. The most common metric to measure the performance of a cache system is the Cache Hit Ratio, which is the ratio of the number of cache hits to the number of lookups, usually expressed as a percentage. Depending on the nature of the cache, the expected hit ratio can vary from $60\%$ to greater than $99\%$.

\subsection{Cache Data Structure}\label{seg:cacheStructure}
With the location of the cache system decided, the next problem is the data structure of the cache system. As discussed in Section\ref{subsec:parallelism:data}, the basic unit in the system is a single column in a chunk; therefore, we chose this same basic unit to represent the cache unit.

\begin{figure}[htbp]
	\centering
	\includegraphics[width=0.85\textwidth]{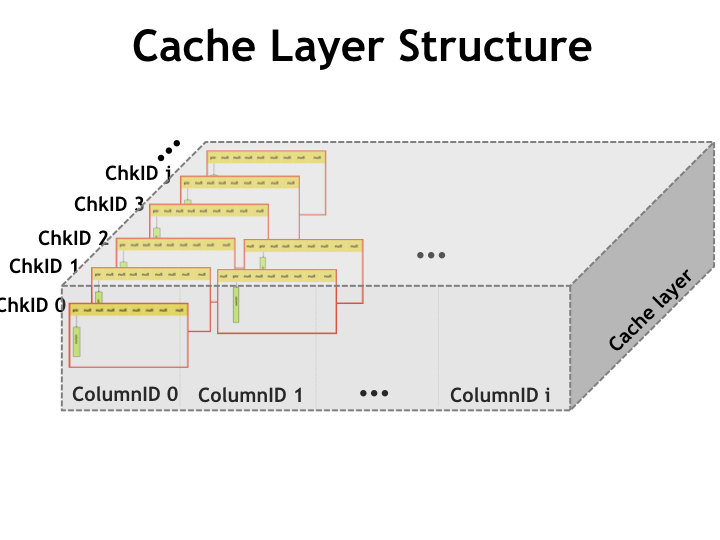}
	\caption{Structure of Cache Data.}
	\label{fig:cacheStructure}
\end{figure}
Commonly, in association with the array of buffer pool frames is a hash table that maps the address currently held in memory to its location in the frame table, the location for that unit on the backing disk storage, and some metadata pertaining to the unit. We also implemented the hash table as the component in our cache system. The hash table key is in the combination form $<ChunkId, ColumnId>$. The value in the hash table is a single column in a chunk instead of a single column itself. Figure\ref{fig:cacheStructure} describes the data structure of our cache system. Each chunk contains an array of pointers used to connect the concrete columns. The structure of the chunks is the same; therefore, when the query requests a chunk, the system can easily merge the $<ChunkId, ColumnId>$ items.

\subsection{Eviction Algorithm}

The ABM and PBM\cite{cooperativescan}work more efficiently than the LRU and LFU for scan-based operations. However, they are not efficient for our storage manager. One reason for this is the complexity, particularly due to its interaction with concurrent updates, data reorganization, and parallel query execution. We avoided additional computation except for the query process and combined the merits of the push- and pull-based models to maximize I/O performance.

In the storage manager, the cache system is referred to as a window-based push model cache (WPC). The cache system is located in the \textbf{TableScanner}. Additionally, a \textbf{Request List} (denoted as RList), lies in the middle and contains a list of chunk IDs indicating which data chunk should be read for query processing. This list is generated by a query optimizer that analyzes the query filter with stored data. The order of the data involved is also important. The paper~\cite{smoothScan} proposes a dynamic mechanism to efficiently generate the list based on data distribution.

Although the data structure of the RList is a simple linked list, it is used by the table scanner to control the chunk read for single or multiple queries. Before describing the WPC in the detail, we need to introduce other core structures that play an important role in the WPC. Section~\ref{seg:cacheStructure} states that the cache data unit is a single column in a chunk. Therefore, the key to defining a unit in the WPC is a pair composed of the chunk and column IDs. After the execution plan is generated for any query, its relevant data is defined by an array of keys (ChunkID and ColumnID).

%
The first step is to check whether the data chunk can be found directly from the cache, meaning that all the chunk and column ID keys are in the cache system. If any key is missing, the system must read it from the disk. To minimize communication costs, the system designs the request list to delegate the table scanner and chunk reader as the producer and consumer of the request list, showing in the middle. The chunk reader takes the chunk IDs from the front of the list one by one for reading. When a chunk is read done from the disk, the chunk reader simultaneously returns it to the table scanner and takes another chunk ID from the request list. When the table scanner receives a chunk from the chunk reader, it attempts to push the chunk into the cache to avoid I/O for the same chunk in the future. Instead of pushing the chunk directly in, the data is divided into multiple small chunks, each containing a single column. For $Q_{1}$ in Example\ref{fun:queries}, the data chunk has two columns, \textit{quantity} and \textit{tax}, that are split into two small chunks before being stored in the cache. Algorithm\ref{alg:pushToCache} depicts the method for caching the data.

\begin{algorithm}[htbp]
	\caption{Put\_Data\_Into\_Cache}\label{alg:pushToCache}
	\begin{algorithmic}[1]	
		\REQUIRE data Cache(DC), Global Status Graph(GSG) and $Read Chunk(RC)$
		\IF{(GSG.getStat(RC) == 0)}
			\STATE\label{A1:l1} $DROP(RC)$;
			\STATE\label{A1:l2} $RETURN$;
		\ENDIF
		
		\IF{($DC.count(RC) > 0$)} 
			\STATE\label{A1:l3} $GSG.update(RC)$;
		\ELSE
			\STATE\label{A1:l4} $DC.insert(RC)$;
			\STATE\label{A1:l5} $GSG.insert(RC)$		
		\ENDIF		
	\end{algorithmic}
\end{algorithm}
The cache system is eager to cache as much data as possible, but if the total size of the cached data reaches the maximal capacity of the system, the system is required to evict lower priority data. We implemented a structure to record the relation between the work unit and all running queries. We designed a Global Status Graph (GSG) that is used to track the data status. The data status contains four fields; the chunk and column IDs compose the key of data status that could define a unique data unit in the cache. To compare priorities between different data units, the data status contains two additional fields, an In-Use Query Number\textbf{(IQN)} and Related Query Number\textbf{(RQN)} represented in Figure~\ref{fig:globalStatus}. The RQN indicates the number of queries that include this chunk in their request list, and the IQN indicates the number of different queries whose request window contains the data unit. Cache units with a high IQN have priority over those with a lower IQN. If the IQN of two data units is the same, the unit with the higher RQN is assigned the higher priority.

\begin{figure}[htbp]
	\centering
	\includegraphics[width=1.0\textwidth]{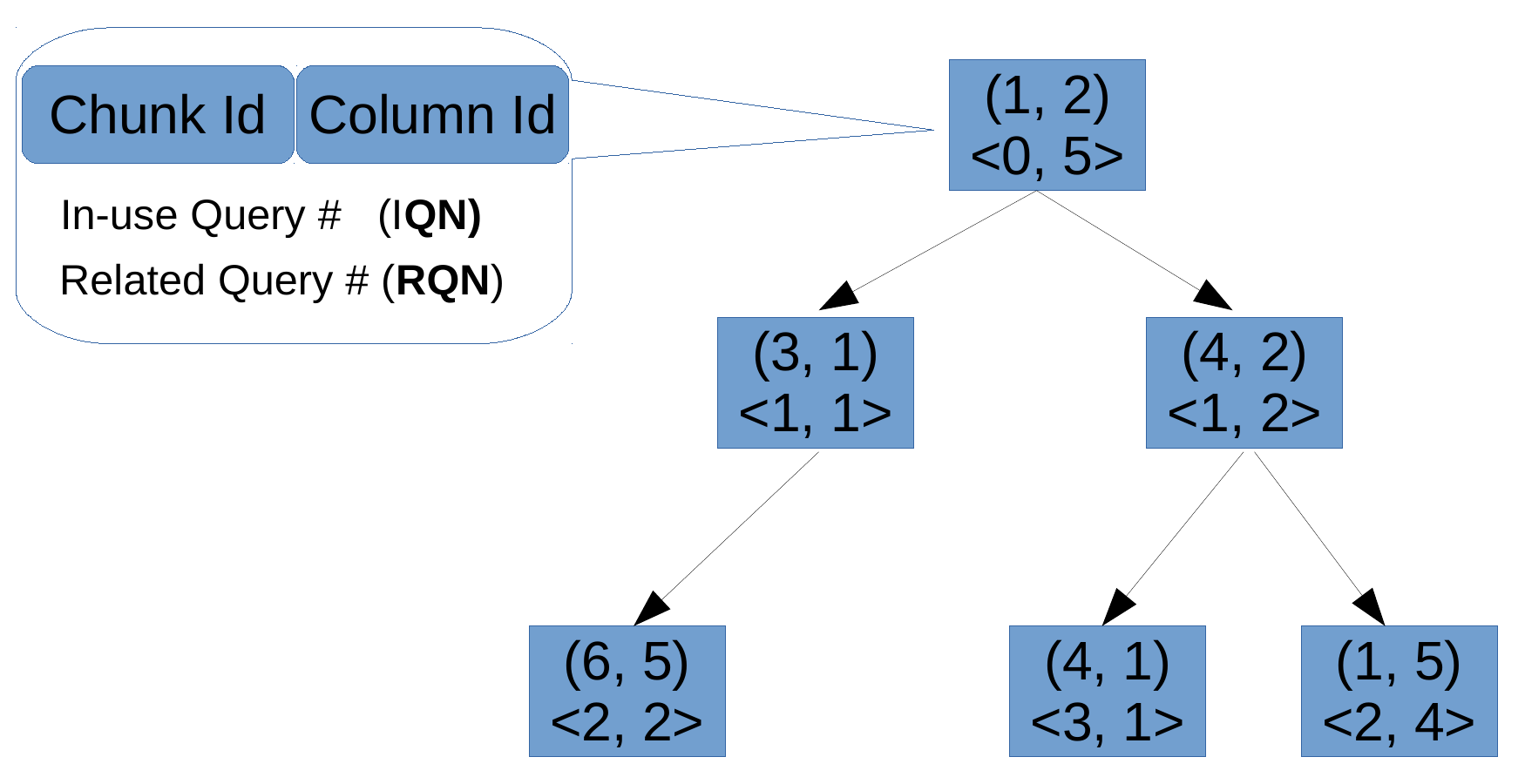}
	\caption{Global Status Graph.}
	\label{fig:globalStatus}
\end{figure}

The \textbf{GSG} tracks the status of the data units in the cache. Its primary task is to select the data unit in the cache with the lowest priority at all times. The GSG utilizes a minimal heap structure to connect the data units; the top of the heap is the unit with the lowest priority. As demonstrated in Figure~\ref{fig:globalStatus}, the top unit of the GSG is the unit with the key (1,2), which indicates $column_{2}$ in $chunk_{1}$. Although the top unit has the highest RQN value (5), its IQN is the lowest, indicating that this unit is not in the query windows of any currently running queries.

\begin{algorithm}[htbp]
	\caption{Get\_Next\_Candidate}\label{alg:getNextCand}
	\begin{algorithmic}[1]	
		\REQUIRE Global Status Graph (GSG)
		\IF{($GSG.size() > 0$)} 
		\STATE\label{A2:l1} $candiate = GSG.getNext()$;
		\STATE\label{A2:l2} $GSG.erase(candiate)$		
		\ENDIF		
	\end{algorithmic}
\end{algorithm}

During multi-queries execution, when the table scanner receives a message indicating the chunk reader has completed the read, the request window moves one step forward and a new chunk ID is included and pushed into the request list -- Chunk 5 for $Q_{1}$. Once Chunk 2 is read back from the disk, the table scanner will receive the notification from the chunk reader and then send the chunk to all the queries belonging to its IQN value. Next, the data chunk’s IQN and RQN decreases respectively. Whenever the storage manager plans to cache the data chunk, it utilizes Algorithm\ref{alg:getNextCand} to locate the candidate from the cache for eviction.
\end{large}

\pagebreak

\chapter{Experiment}\label{ch:experiment}
\begin{large}
The objective of the experimental evaluation is to investigate the performance of the \texttt{Storage Manager} across a variety of workloads, including a single query and batch queries. Additionally, we ran extensive experiments with many configuration parameters. The designed experiments answer the following topics:

\begin{itemize}
	\item •	How the NVMe SSD improves the performance of the \texttt{Storage Manager}.
	\item •	How the parameter configuration affects the performance of the \texttt{Storage Manager} performance.
	\item •	How the \texttt{Storage Manager} behaves compared to a traditional storage manager engine
\end{itemize}

\paragraph*{Implementation}
The \texttt{Storage Manager} was implemented as a \texttt{C++} prototype. Each standalone thread, as well as the workers, were implemented as \texttt{pthread} instances. The code contains special function calls to harness detailed profiling data. In the experiments, we implemented the \texttt{Storage Manager} with a cutting-edge multi-thread database system~\cite{datapath, glade,extascid} demonstrated to effectuate hyper-efficient implementation. In this experiment, we implemented both the CS and HighTh models; the main difference between the two models is the query access plan schedule. This guarantees that the following query performance comparison is only related to the \texttt{Storage Manager}. In addition, the CS model adopts the pull-based model, which indicates the request window size should always be 1.

\paragraph*{System}
We perform the experiments on a dual-socket server with two 14-core Intel Xeon E5-2660 v4 CPUs (56 threads overall, 256 GB memory) and eight Seagate 2TB Enterprise Capacity 2.5 HDD V.3 and two Intel 800GB DC P3700 Series HET-MLC. Each processor has 35 MB L3 cache while each core has 32 KB L1 and 256 KB L2 local caches. According to \texttt{hdparm}, the cached and buffered read rates in SSD are 10 GB/second and 3.3 GB/second. The buffered read rates from HD is 900 MB/second. Ubuntu 16.04.6 LTS $64$-bit with Linux ubuntu16 4.4.0-98-generic is the operating system.

\paragraph*{Methodology}
We performed every experiment a minimum of three times and reported the average value as the result. If the experiment consisted of a single query, we enforced the data to be read from the disk by cleaning the file system buffers before execution. In experiments over a sequence of queries, the buffers were cleaned only before the first query. Thus, the second and subsequent queries could access cached data.

\paragraph*{Data.}

In the experiment, we ran a batch of queries on TPC-H, which is a benchmark that simulates a decision support system or business intelligence database environment. The performance of a system is measured when the system is tasked with providing answers for business analyses on a dataset. Therefore, it is popular for comparing database vendors. The TPC-H database components were defined to consist of eight separate tables. LINEITEM was the most important table, taking more than $80\%$ of the data set. To control the size of data set, TPC-H datasets were scaled by a scale factor, which is a multiplier of the number of rows in each table. A scale factor of $1$ represents a dataset of roughly $1$ GB. The tables were populated using DBGEN, a data generator provided in the TPC-H package to populate the database tables with different amounts of synthetic data. In this experiment, we chose a scale factor of $40$; the size of the generated file was approximately 40 GB.

\paragraph*{Queries.}
The query used throughout experiments has the form \texttt{SELECT FUN($C_{i}$) FROM LINEITEM WHERE $C_{i}$ $\geq \alpha_{i}$}, where $C_{i}$ is in the columns list of the table LINEITEM. The columns of each query is generated by randomly picking up a subset of table columns at the beginning. The $\alpha_{i}$ defines the user-defined filter according to column data type, the value is produced by the standard $normal\_distribution$ function to generate a random value between its min and max value. This type of query reads tuples from the data and process those tuples by a user-defined function, which is super simple function, such as $SUM$ or $AVG$. The workload in the experiment consists of 10 sets of batches, each of them contains 64 queries which have the same query type as described at the beginning of the paragraph. In the following experiments, we comprehensively compare the our storage manager with the cooperative-scan implementation with different configurations, such as storage material(HD vs SSD), chunk size, cache size and request query window size, etc.

\begin{figure}[htbp]
	\centering
	\includegraphics[width=0.95\textwidth]{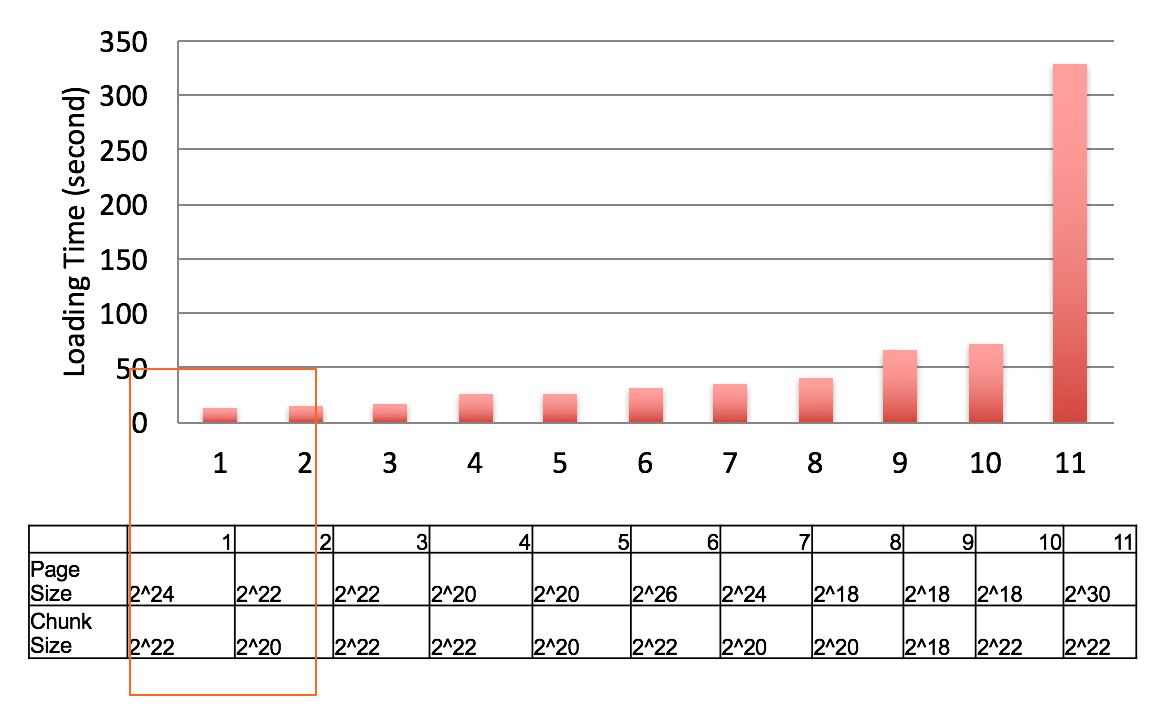}
	\caption{Chunk Size Comparison.}
	\label{fig:pruning}
\end{figure}

\section{Parameter evaluation.} 
The system has two configurable parameters, \textbf{Page Size} and \textbf{Chunk Size}. The page size indicates the basic storage unit in the system, and the chunk size indicates how to split the data based on the number of tuples. As the chunk size increases, the number of chunks decreases. The page size determines the number of I/O requests per chunk. The first experiment evaluated the performance of the queries’ workload with different combinations of chunk and page sizes. In this experiment, we measured the ETL process as the benchmark because the standard ETL process includes the ETL stages and is therefore able to evaluate the performance comprehensively. The results in Figure~\ref{fig:pruning} indicate that the optimal chunk and page sizes are approximately $2^{22}$ and $2^{24}$ bytes, respectively.
 
\begin{figure}[htbp]
	\centering
	\includegraphics[width=0.95\textwidth]{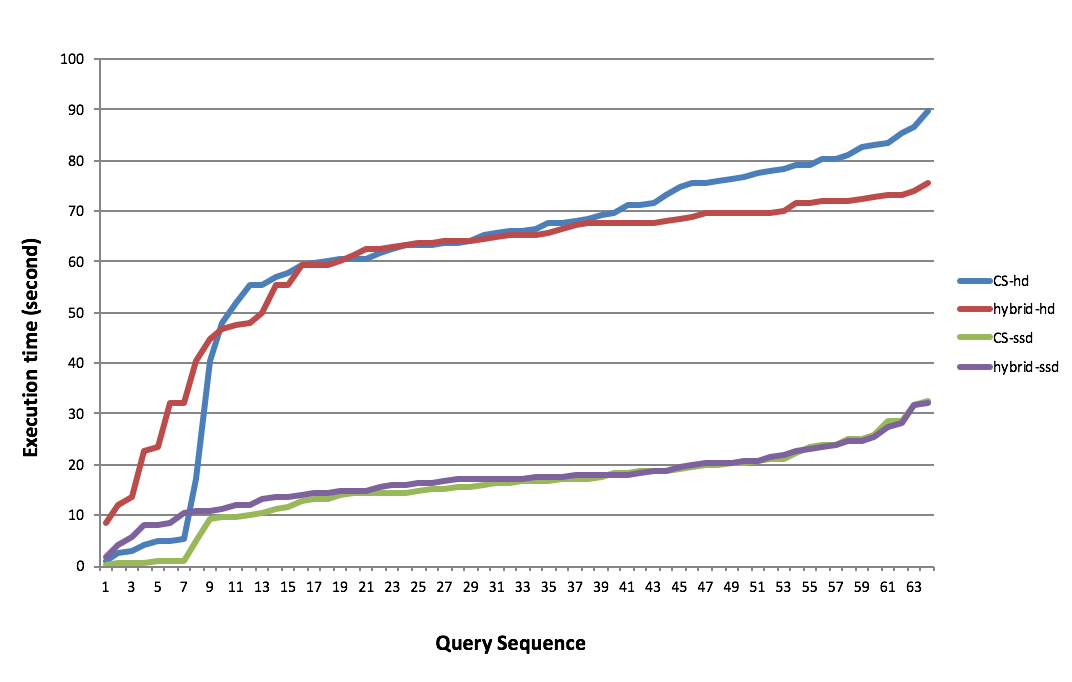}
	\caption{Query Performance Comparison.}
	\label{fig:queryperformance}
\end{figure}

\section{Query Performance Comparison.} 
The experiment depicted in Figure~\ref{fig:queryperformance} compared the performance of four scenarios: the CS with an unlimited cache storage manager (CS) versus an optimized version (HighTh model) using an HDD and SSD, respectively. The experiment sought to execute a batch of queries (64 in total) together and measure the accumulative execution time after each query was complete. Figure~\ref{fig:queryperformance} illustrates the overall execution time after the $i$ queries were finished, where $i$ goes from 1 to 64. The horizontal axis represents the query ID in the workload, and the vertical axis is the accumulative execution time after the  $i-th$ query is complete.

As expected, queries running on the SSD were significantly faster than those running on the HDD. The high I/O throughput of the NVMe SSD shows that the total execution time on the SSD dropped from 80 to 30 seconds, compared to running on the HDD.

On both the HDD and SSD, the CS ran faster at the beginning for the small group of queries in front. The reason is that the CS storage manager always reads only the required data; therefore, the read time is minimal. Compared to the CS, in the new HighTh model, the storage manager focuses on the query plan for the entire batch of queries. Data chunks in the running queries’ request windows are merged to maximize I/O performance. This process may increase the amount of data to be prepared for the initial queries, which would explain why the few queries running in the HighTh model were comparatively slower. However, as more and more queries were finished, the HighTh model performed better. After many queries were completed, some of the data was in the cache, so the following queries could receive some data directly from the cache. The chunk merge strategy could prefetch data efficiently. Compared to the CS, the new HighTh model storage manager optimized the query access plan at the upper level to minimize the global execution time by maximizing the I/O throughput as much as possible. 

The experiment demonstrates the merit of the new storage manager, especially on the HDD. Since the I/O speed of the HDD is relatively slow, the queries in the experiments running on the HDD were usually I/O bound. Optimizing the query access plan can efficiently increase the I/O throughput of the HDD, which is why the total execution time of the HighTh model was less than the CS. In the SSD environment, the execution time followed a similar pattern. Due to the high throughput of the NVMe SSD, the query execution was normally CPU bound, but even in this situation, the HighTh model performance was not inferior to that of the CS.

\section{Query running on Hard Disk.} 
\begin{figure}[htbp]
	\centering
	\includegraphics[width=1.05\textwidth]{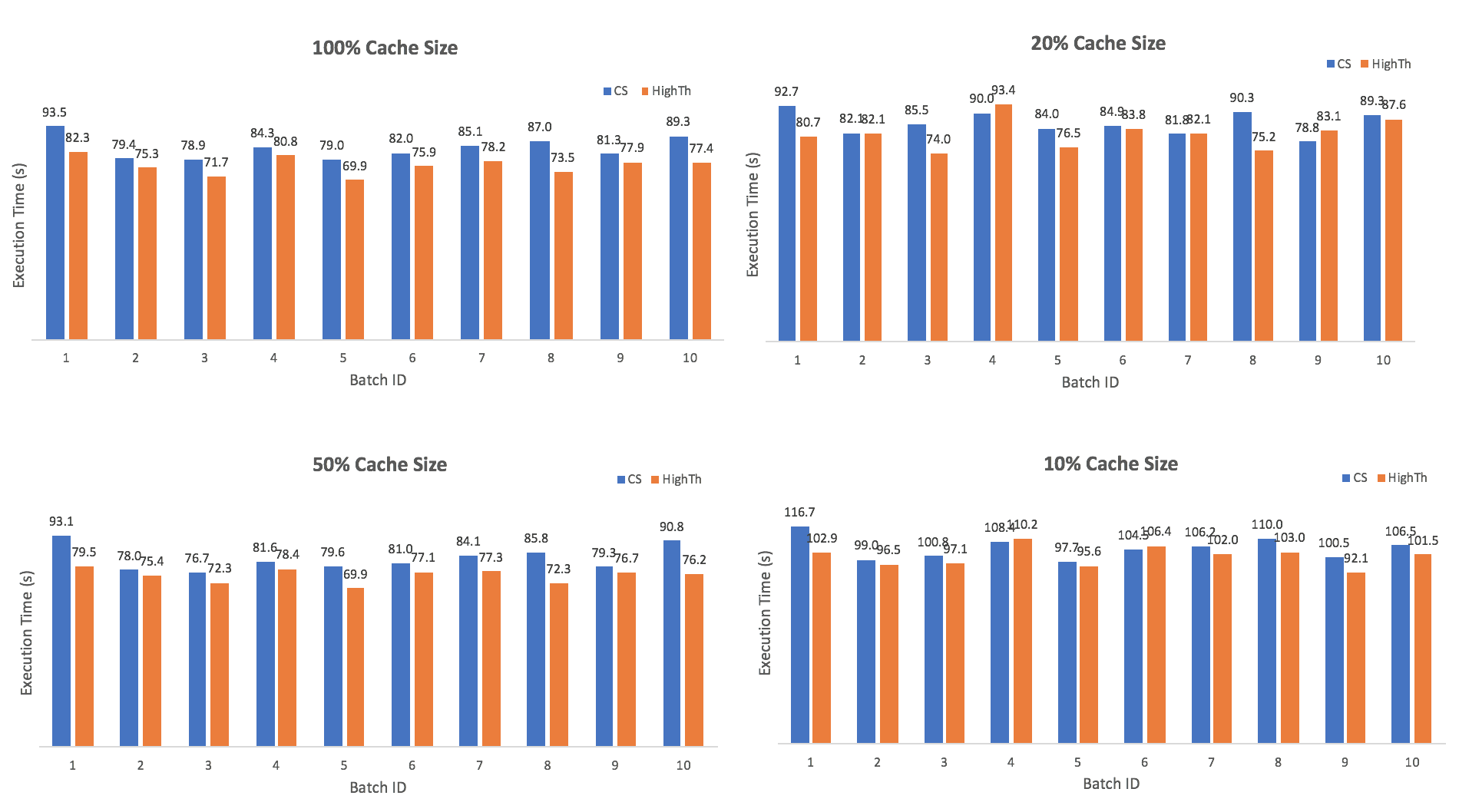}
	\caption{Query Execution on Hard Disk.}
	\label{fig:hdfull}
\end{figure}
To evaluate the performance of the HighTh model when processing batch queries, we first generated a series of workloads to represent the batch queries. For each experiment, we ran 10 batches of queries independently. For each batch, there were 64 queries in the \textit{Lineitem} table. The main performance metric we measured was the execution time and I/O number. We also measured the power of the eviction algorithm when the size of the cache buffer met the limitation. We allocated the size of cache with $10\%$, $20\%$, $50\%$ and an unlimited percent of the total data size.

Figure~\ref{fig:hdfull} depicts the total execution time of the CS and the HighTh model. The horizontal axis represents the query batch ID in the workload, and the vertical axis shows the execution time for the $i-th$ batch. The results indicate that as the cache size decreased from unlimited to $10\%$, the queries’ execution time increased accordingly for both models. Due to the cache limitation, the storage manager could not hold the entirety of the data in memory all the time, so it had to regenerate some data chunks from the disk multiple times, which resulted in a longer execution time. In most cases, regardless of whether or not the cache size was limit, the HighTh model consistently outperformed the CS with a shorter execution time. Since the I/O speed is slow in the HDD and we did not set the limitation of the CPU in these experiments, all queries running on the HDD were considered to be I/O bound, which means that the majority of the execution time was relay on the data read time from the disk. The main difference between the two models is that the HighTh model generated the data access plan of the entire workload, 64 queries in our example, instead of considering only the single query. The new storage manager aims to maximize the I/O access between multiple queries and priorities the data according to the \textbf{IQN} and \textbf{RQN} described in Section\ref{sec:bufferCache}.  

\begin{figure}[htbp]
	\centering
	\includegraphics[width=1.0\textwidth]{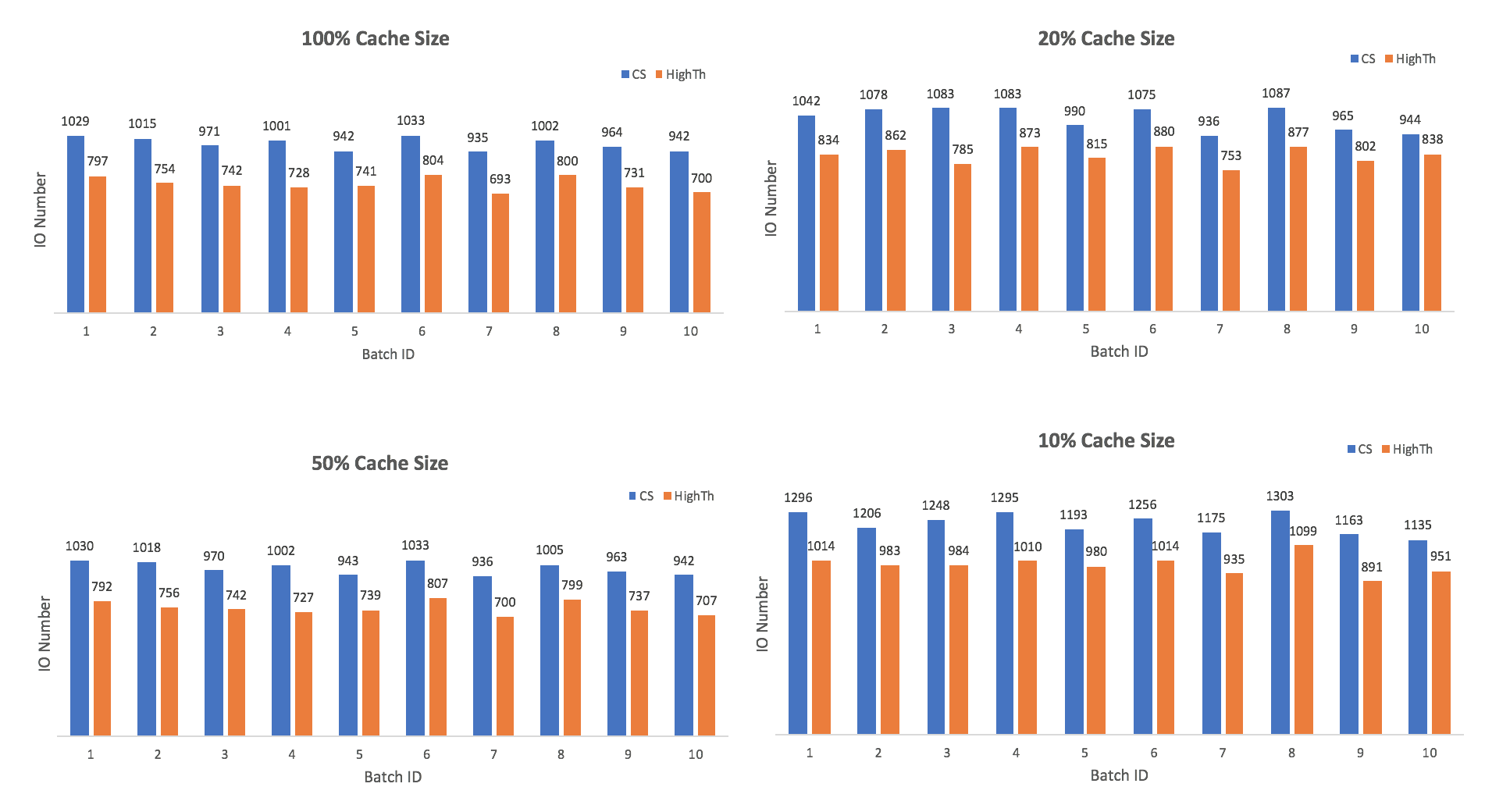}
	\caption{I/O requests on Hard Disk.}
	\label{fig:IO-hd}
\end{figure}

When evaluating the HighTh model’s impact on I/O performance, we measured the read requests sent to the disk. Figure\ref{fig:IO-hd} depicts the comparison between the CS and the HighTh model with different cache sizes. The results indicate that the I/O number increased when the cache size decreased. As previously stated, when the buffer is not large enough to cache all the data, the storage manager evicts some data to make space. The smaller the cache size, the more the eviction occurs. Even when the cache size was  $10\%$, the I/O number of the HighTh model was consistently lower than the CS model. This is because the HighTh model merges multiple read requests according to the same chunk into a single larger request, minimizing the total number of read requests sent to the disk. Moreover, since the data of the same chunk is located nearby, the merging improves both the read throughput and efficiency. Therefore, like the CS model, the HighTh model evicts the less hot data and serves the current query. In addition, we were able to change the read order to satisfy the current request while simultaneously improving the I/O number.

\begin{figure}[htbp]
	\centering
	\includegraphics[width=1.0\textwidth]{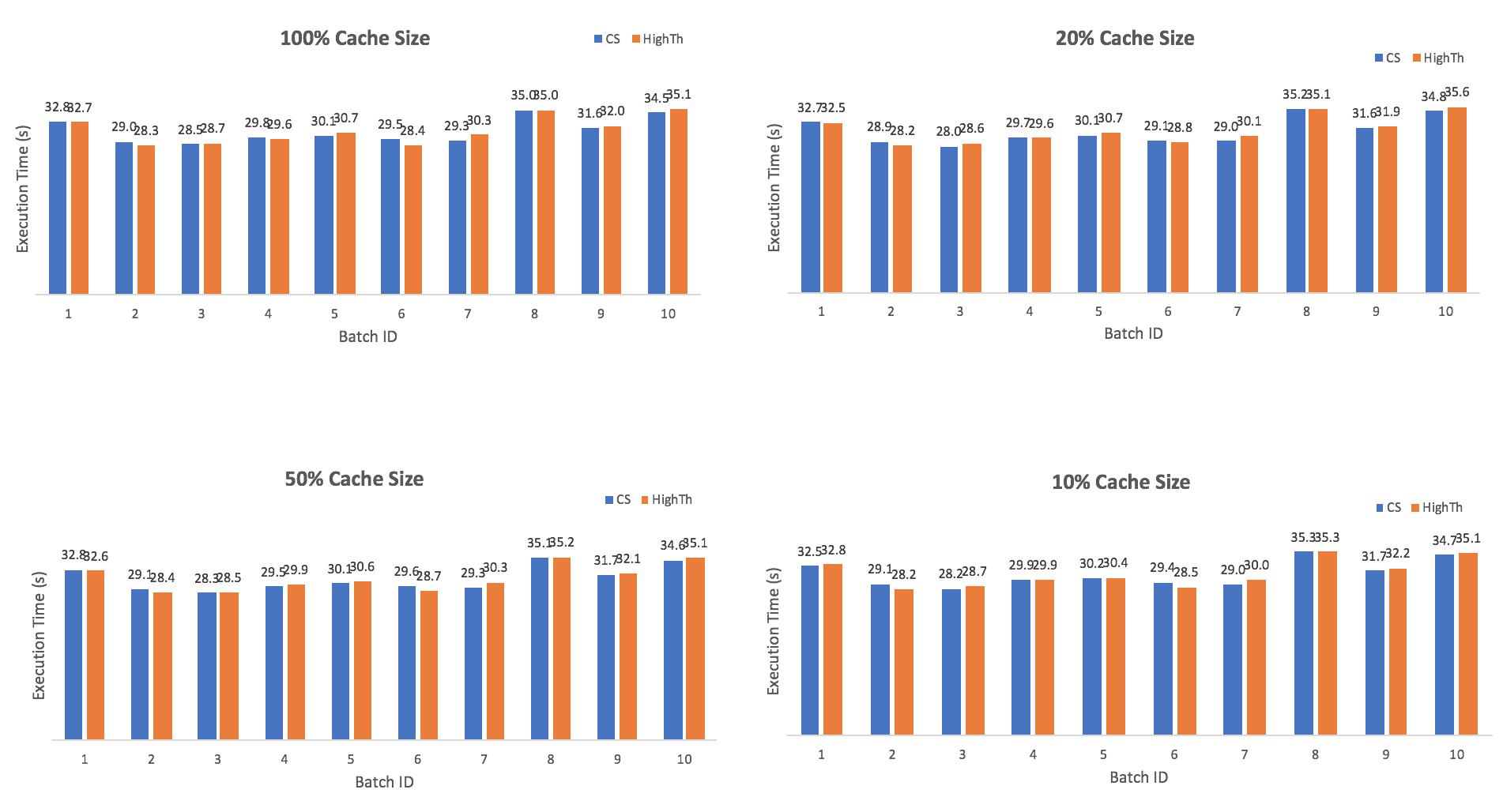}
	\caption{Pull-based Query Execution on SSD.}
	\label{fig:ssdfull}
\end{figure}

\section{Query running on SSD.} 
To fully understand the performance of the new storage manager, we ran the same experiments on the NVMe SSD. Figure~\ref{fig:ssdfull} depicts the performance of the same workloads as the HDD on the SSD. As expected, due to the high performance of the SSD, the execution time was much faster than that of the HDD. However, unlike the performance comparison on the HDD, both models achieved a similar execution time. The primary reason for this result is that the read time is much faster in the SSD compared to the HDD, and we limited the CPU number up to 25, so the processing in this experiment was considered to be CPU bound for both of the models. The execution time was mainly decided by the computation cost rather than the read time.

We ran the above experiments with unlimited computation resources, which converted the queries from CPU bound to I/O bound. As understood from Figure~\ref{fig:ssdfull2}, the execution time of the CS model was not affected by the computation resources. However, the HighTh model improved the execution time around 10 seconds. In the results, the new storage manager constantly decreased the execution time with different cache size strategies. This was also due to merging the requests from multiple queries and fetching the data chunks in advance.

\begin{figure}[htbp]
	\centering
	\includegraphics[width=1.0\textwidth]{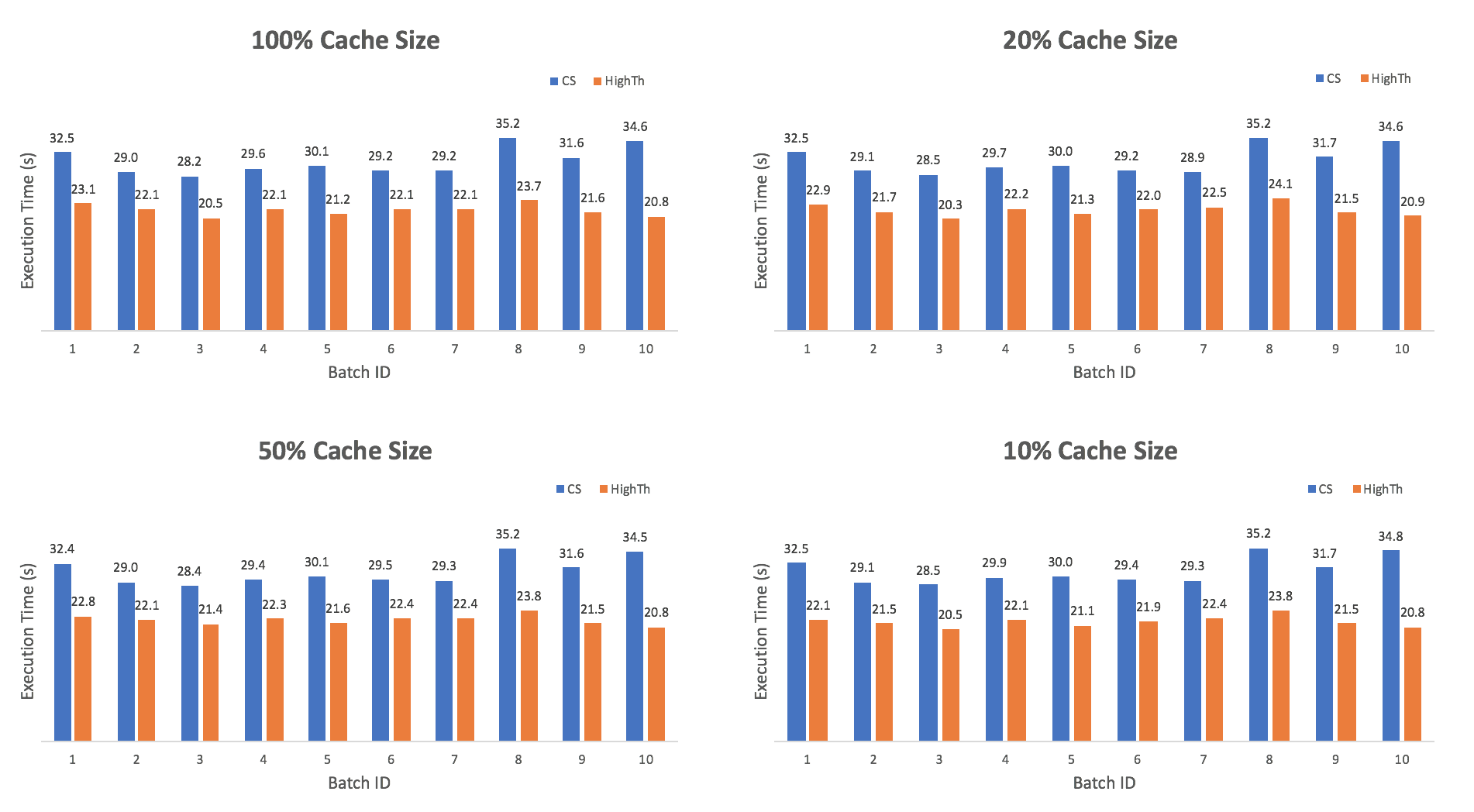}
	\caption{Push-based Query Execution on SSD. }
	\label{fig:ssdfull2}
\end{figure}
As discussed in Section\ref{sec:bufferCache}, the HighTh storage manager can dramatically decrease the I/O number. Figure~\ref{fig:IO} depicts the I/O process comparison between the CS and the HighTh model in the SSD. Similar to the HDD results, when the size of cache shrank, the number of I/O requests sent to the disk increased. However, the number of I/Os for the HighTh model was much fewer than in the CS model, about $70\%$. This confirms the results from the HDD that the storage manager generates an optimized data access plan based on the entire workload, which reorganizes the data chunks’ access order to maximize access sharing. Multiple small read requests from different queries are merged together, therefore reducing both the execution time and I/O number.

\begin{figure}[htbp]
	\centering
	\includegraphics[width=1.0\textwidth]{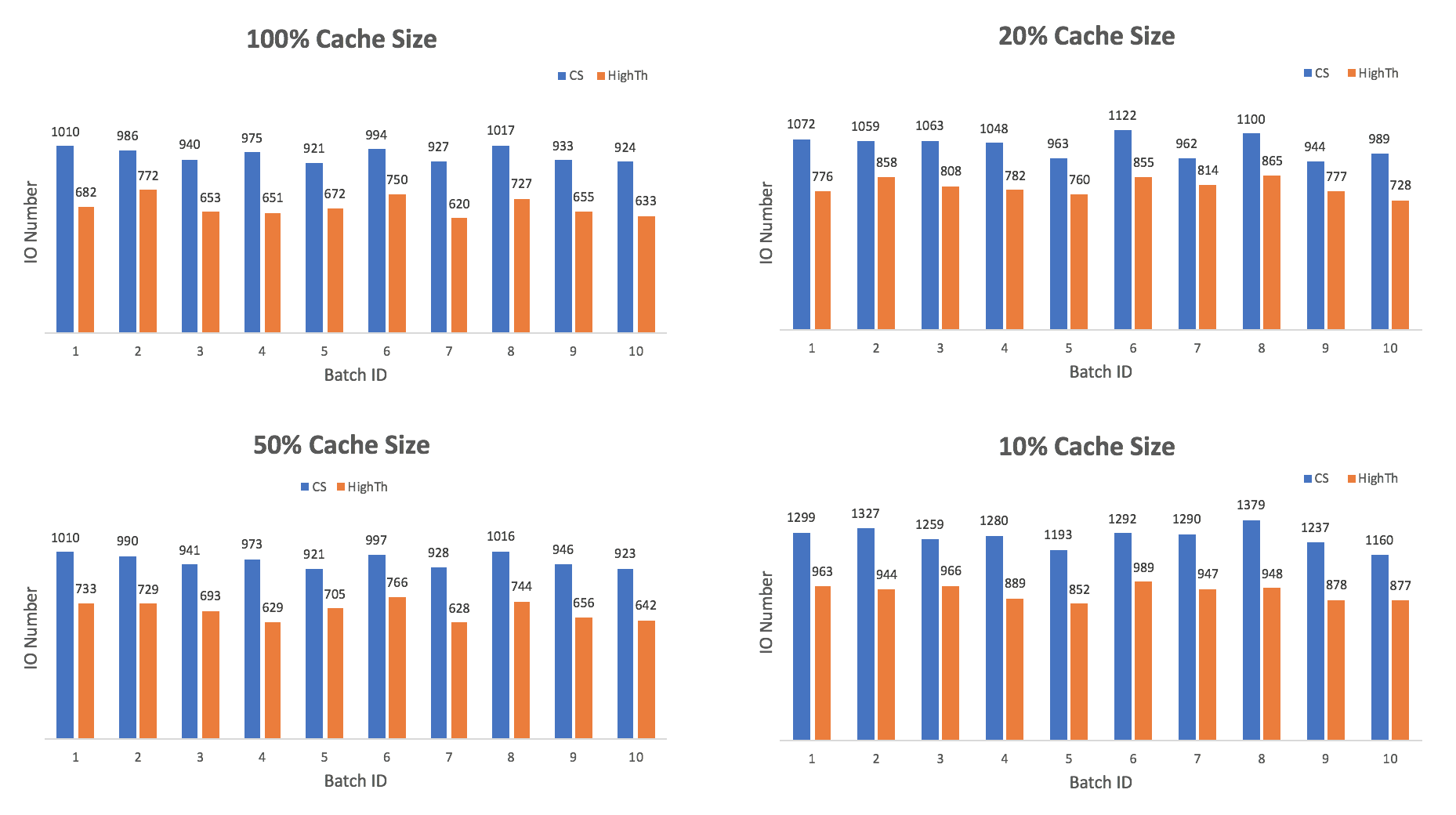}
	\caption{I/O requests on SSD.}
	\label{fig:IO}
\end{figure}

The results in Figure~\ref{fig:avgTimeIO} indicate a significant improvement in the execution time when applying the SSD for both models. In the HDD, the execution time increased when the cache size decreased, while the execution time in the SSD stayed relatively the same. The execution time of the HighTh model was consistently lower than that of the CS model in both the HDD and SSD among different cache sizes. When examining the I/O numbers, we discovered that the HighTh model always performed better than the CS model with fewer I/O requests.

\begin{figure}[htbp]
	\centering
	\includegraphics[width=1.0\textwidth]{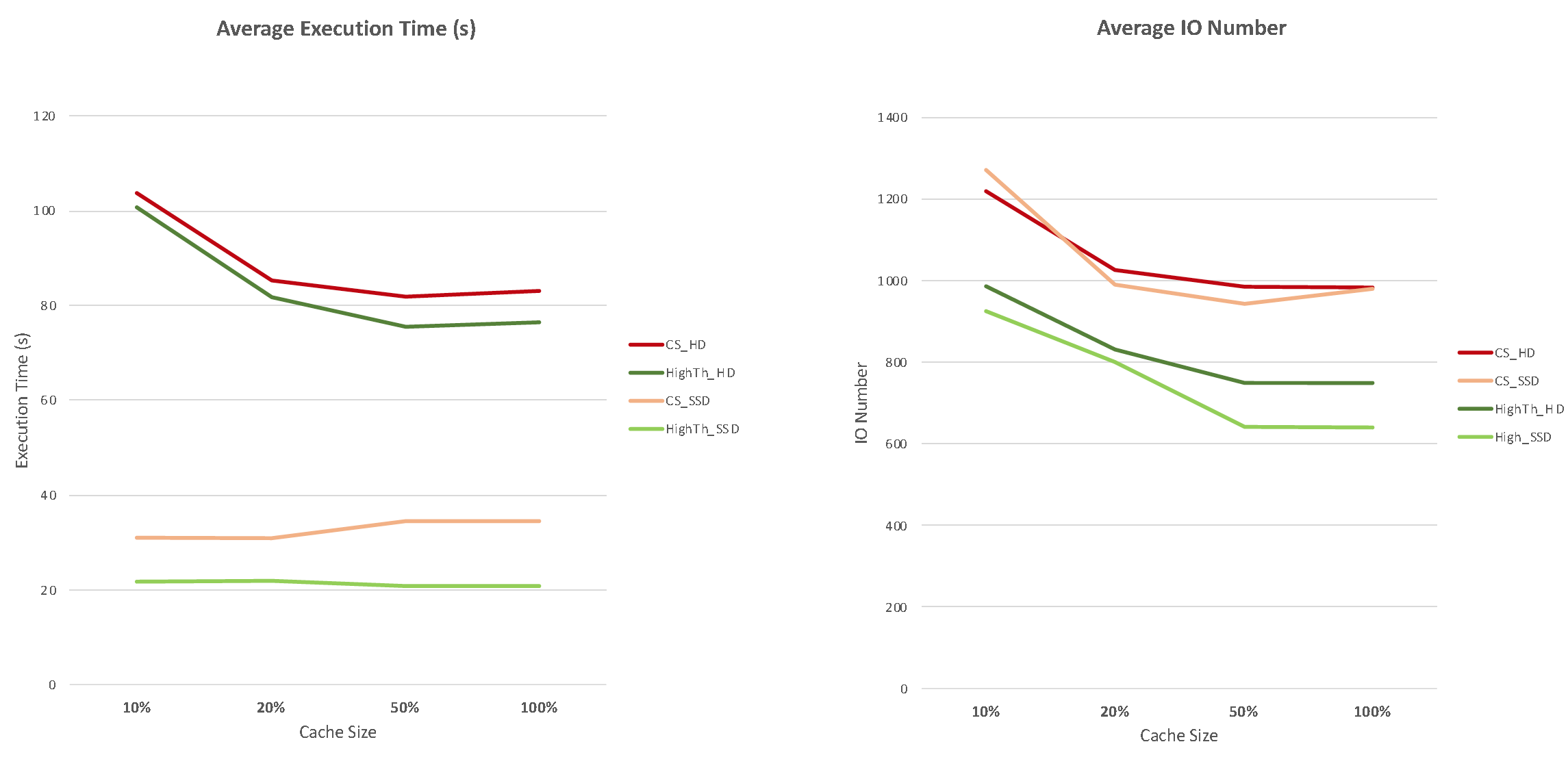}
	\caption{Comparison bewteen CS and HighTH in both HD and SSD.}
	\label{fig:avgTimeIO}
\end{figure}

We also evaluated the prefetch window size as another parameter in our experiments. We modified the window size among $1$, $5$, $10$, $15$ and $30$  to run the same experiments. Since the CS model does not have a merge mechanism, we treated it as a window size equal to 1. We would like to see whether the window size affected the performance. In order to understand the characteristics of the two models, we also compared the execution time versus the read time. The execution time here means the total time from the start until the end of the processing. The read time is counted from the start of the processing and ends after the last read request is sent to the disk. If the subsequent queries could find the data in the cache, no further read requests were sent. We measured the difference between the query execution time and read time. The difference represents the improvement of the model producing the necessary data. The earlier, the better.

Since the result of the experiments on the SSD and HDD produce the same pattern, we chose to illustrate the details for the experiments run on the SSD. Figure~\ref{fig:} contains the query execution and final chunk read times, and the figures in the lower portion show the difference. When the cache was very limited, such as at $10\%$, the CS and the HighTh model with a small window size, such as W = 5, took longer to process the query; moreover, the difference between the query execution and the read times was minimal. 

However, when the window size increased, the performance of the HighTh model improved. Both the query execution and final chunk read times decreased. As the window size increased, the difference between the query execution and final read times also increased. The main reason for this increase is that with a large window size, the HighTh model has more opportunities to optimize the access plan for batch queries. Even with a small cache size, the query access plan can be very efficient.  

\begin{figure}[htbp]
	\centering
	\includegraphics[width=1.0\textwidth]{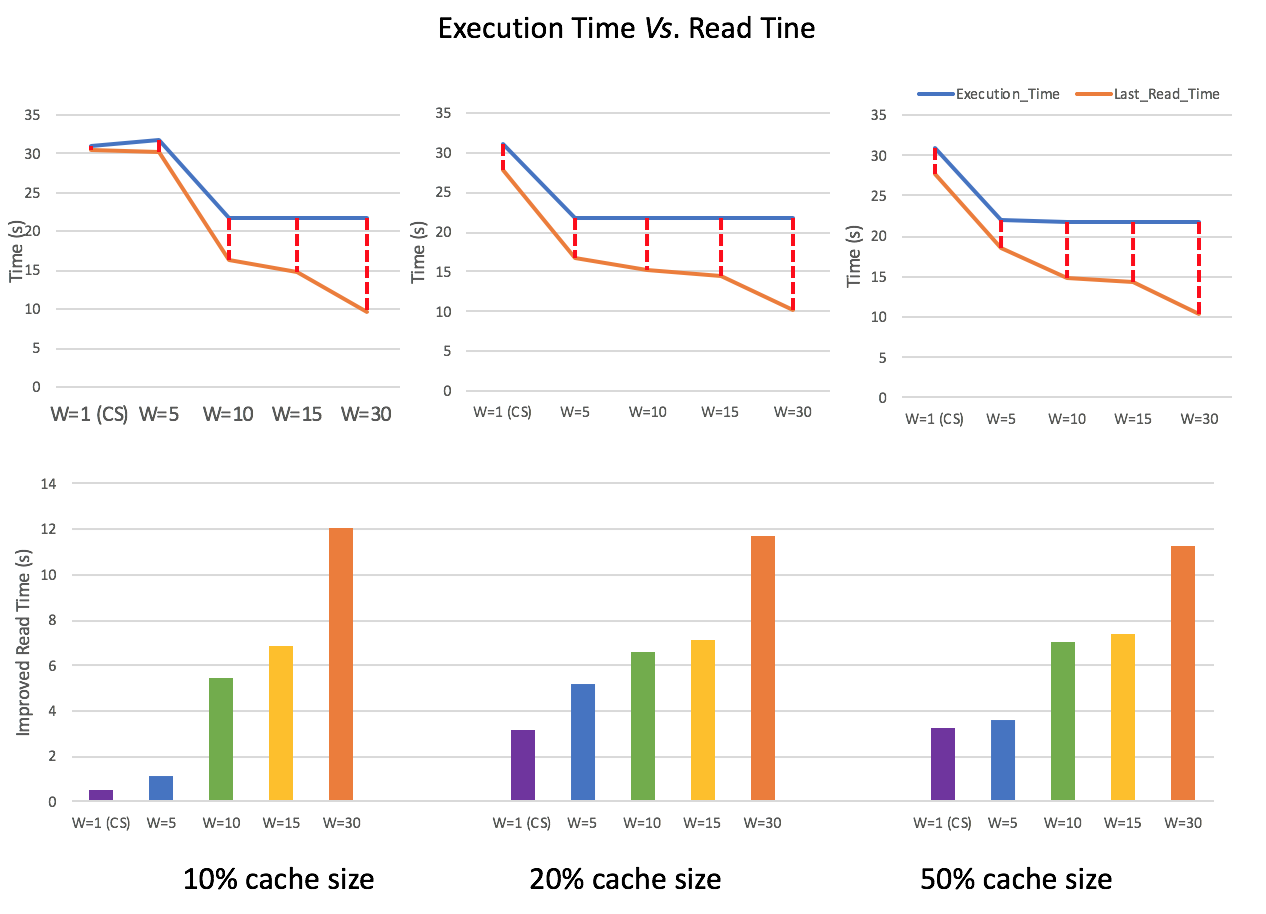}
	\caption{Difference between execution time and data reading time.}
	\label{fig:last}
\end{figure}

Finally, we also compared the I/O numbers of the query workloads with different window sizes. Figure\ref{fig:winio} provides the average I/O numbers calculated from the 10 batches. The I/O number from the CS model was the highest. When the window size increased from 5 to 30, the trend of the I/O number in the HighTh model became increasingly flat, which means the cache size did not affect the I/O request numbers because the larger the window size, the more opportunities there are to merge the small requests. Moreover, the HighTh model has more chances to reorder the requests among multiple queries to generate an optimized access plan. In an ideal situation, the storage manager would only need to read the data chunk once, so there would be no need cache. In the experiment, a window size of 30 was the optimal condition since the total number of required chunks per query was around 30. The results also confirm this, showing that the I/O number remains the same among the different cache sizes.

\begin{figure}[htbp]
	\centering
	\includegraphics[width=.9\textwidth]{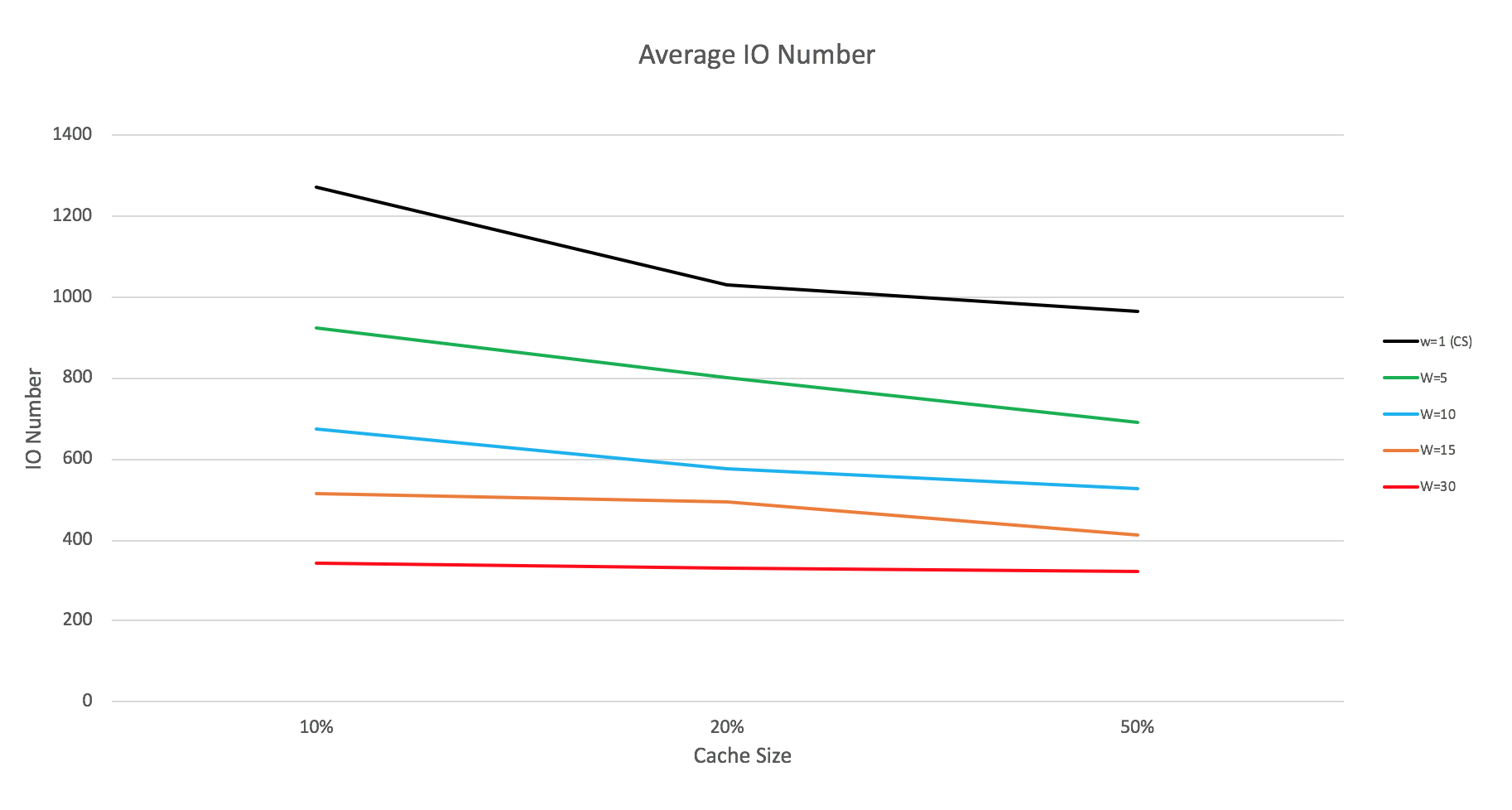}
	\caption{I/O number with different window size on SSD.}
	\label{fig:winio}
\end{figure}

From the above discussion, we can conclude that the HighTh storage manager outperforms the CS in all configurations for different batch query workloads. Although a few queries’ access plan could be changed to make their own execution time relatively slow, the execution for the entire batch is still more efficient, both in HDD and SSD environments. Therefore, the HighTh model processes batches of queries more efficiently.

\end{large}
\pagebreak

\chapter{Conclusions}\label{ch:conclusion}
\begin{large}
In this thesis, we analyze the storage manager on database systems, improving the performance issues of the buffer cache layer, data access layer and the disk layout layer. This includes analyzing the characteristics of workloads running on database system, identifying the performance In this thesis, we analyzed the storage manager on database systems and improved performance issues in the buffer cache, data access, and disk layout layers. The scope of this project included analyzing the characteristics of workloads running on database systems, identifying the performance bottleneck of the storage manager in different layers, and designing and evaluating new components and algorithms capable of improving performance. We designed and implemented a high-performance storage manager composed of multiple components running in parallel. We also discovered an optimal method to organize data inside the system. In addition, we designed an efficient cache management component in the storage manager, referred to as a WPC. Finally, we ran extensive experiments to evaluate the performance of the storage manager. The results indicated that the performance of the new storage manager outweighed the standard system in many dimensions.

Hardware performance became faster, especially for the NVMe. In the future, we plan to evaluate the storage manager performance for the new version of the hardware. Additionally, due to the high computation capacity, GPU has been widely used in many systems. However, its memory size is limited and the transfer cost between the host and GPU is very high. Going forward, we could adjust the cache management system to be suitable to compute in GPU. Future studies could also focus on ascertaining an optimized window size for each query according to whether it was CPU- or I/O-bound. Queries could obtain different window sizes under this novel storage management model.

\end{large}

\pagebreak

\bibliographystyle{abbrv}	
\bibliography{Dissertation}

\end{document}